\documentstyle[12pt,amssymb]{article}

\newcommand{\beq}{\begin{equation}}
\newcommand{\eeq}{\end{equation}}
\newcommand{\bea}{\begin{array}}
\newcommand{\eea}{\end{array}}
\newcommand{\bey}{\begin{eqnarray}} 
\newcommand{\pslash}{\not{\hbox{\kern-2.3pt $p$}}}
\newcommand{\pdslash}{\not{\hbox{\kern-2pt $\partial$}}}
\newcommand{\bfx}{{\bf x}}

\newcommand{\0}{{\bf 0}_{4\times 4}}
\newcommand{\bgamma}{{\bf{\gamma}}}
\newcommand{\bGamma}{{\bf\Gamma}}
\newcommand{\psibar}{{\overline{\psi}}}

\newcommand{\Psibar}{{\overline{\Psi}}}
\newcommand{\nubar}{{\overline{\nu}}}

\newcommand{\ubar}{{\overline{u}}}
\newcommand{\vbar}{{\overline{v}}}

\newcommand{\E}{{\epsilon}}
\newcommand{\bsig}{{\mathbf{\sigma}}}

\newcommand{\un}{{\mathbb{I}}}

\newcommand{\bp}{{\bf p}}
\newcommand{\ri}{{\rm i}}
\newcommand{\bP}{{\bf P}}

\newcommand{\pd}{\partial}

\newcommand{\bv}{{\bf v}}
\newcommand{\tr}{{\rm Tr}}
\newcommand{\bfn}{{\mathbf{\nabla}}}
\newcommand{\eey}{\end{eqnarray}}

\def\JPAT#1#2#3#4{#2 #1 {\em J. Phys. A: Math. Theor.} {\bf #3} #4}
\def\FP#1#2#3#4{#2 #1 {\em Fortschr. Phys.} {\bf #3} #4}

\def\AJM#1#2#3#4{#2 #1 {\em Am. J. Math.} {\bf #3} #4}
\def\APNY#1#2#3#4{#2 #1 {\em Ann. Phys. (NY)} {\bf #3} #4}
\def\PTP#1#2#3#4{#2 #1 {\em Prog. Theor. Phys.} {\bf #3} #4}
\def\PR#1#2#3#4{#2 #1 {\em Phys. Rev.} {\bf #3} #4}

\begin{document}

\begin{titlepage}
\vskip 2cm
\begin{center}
{\Large\bf An 8-dimensional realization of the Clifford algebra in the 
 5-dimensional Galilean space-time
\footnote{{\tt masanori@gifu-u.ac.jp,}
 {\tt montigny@phys.ualberta.ca,}  {\tt khanna@phys.ualberta.ca} }} 
\vskip 3cm
{\bf 
M. Kobayashi$^{a,b}$, M. de Montigny$^{a,c}$, F.C. Khanna$^{a,d}$ \\} 
\vskip 5pt
{\sl $^a$Theoretical Physics Institute, University of Alberta\\
 Edmonton, Alberta, Canada T6G 2J1\\}
\vskip 2pt
{\sl $^b$Department of Physics, Gifu University\\
  Gifu, Japan 501-1193\\}
\vskip 2pt
{\sl $^c$Campus Saint-Jean, University of Alberta \\
 Edmonton, Alberta, Canada T6C 4G9\\}
\vskip 2pt
{\sl $^d$TRIUMF, 4004, Wesbrook Mall\\
Vancouver, British Columbia, Canada  V6T 2A3\\}
\vskip 2pt

\end{center}
\vskip .5cm
\rm

\begin{abstract}
We give an 8-dimensional realization of the Clifford algebra in
 the 5-dimensional Galilean space-time by using a dimensional
 reduction from the $(5+1)$ Minkowski space-time to the $(4+1)$
 Minkowski space-time which encompasses the Galilean space-time.
 A set of solutions of the Dirac-type equation in the 5-dimensional
 Galilean space-time is obtained, based on the Pauli
 representation of $8\times 8$ gamma matrices. In order to find
 an explicit solution, we diagonalize the Klein-Gordon divisor by
 using the Galilean boost.
\end{abstract}

\end{titlepage}

\setcounter{footnote}{0} \setcounter{page}{1} \setcounter{section}{0} %
\setcounter{subsection}{0} \setcounter{subsubsection}{0}

%%%%%%%%%%%%%%%%%%

%%%        INTRODUCTION

%%%%%%%%%%%%%%%%%%

\section{Introduction}

Nearly twenty years ago, Takahashi investigated the 
 reduction from a $(4+1)$ Galilean covariant manifold
 to the Newtonian space-time (with 3-dimensional space) as means to build
 non-relativistic many-body theories by starting with Lorentz-like,
 manifestly covariant, equations \cite{takahashi1988}.
 This `Galilean manifold' is actually a $(4+1)$ Minkowski
 space-time with light-cone coordinates, which is reduced to the
 usual Newtonian space-time \cite{omote}. 
 Galilean covariant theories for the Dirac-type fields have been 
 developed by using a 4-dimensional realization of the Clifford
 algebra in a 5-dimensional Galilean space-time \cite{omote}.
 Therein, we have 16 independent components that may be expressed
 as $\gamma_A=\un$, $\gamma_\mu$, $\sigma_{\mu\nu}$,
 with $\mu, \nu = 1,\dots, 5$ \cite{kobayashi}.
 Unfortunately, none of the pseudo-tensor interactions of rank 0, 1 and 2
 can be introduced into 5- (or any odd-) dimensional theories, because
 they admit no `$\gamma^6$ matrix' which corresponds
 to the $\gamma^5$ of the $(3+1)$ Minkowski
 space-time. A 4-dimensional realization of the Clifford
 algebra in the $(4+1)$ Minkowski space-time requires
 $\gamma^5$ as a fourth spatial element of $\gamma_\mu$s.  
 Motivated by this fact, we discuss in this article
 an 8-dimensional realization of the Clifford algebra in the $(4+1)$
 Galilean space-time. 
 Thus our formulation involves two successive
 dimensional reductions: from the $(5+1)$ Minkowski space-time to
 a $(4+1)$ Minkowski space-time, which corresponds to the 5-dimensional
 Galilean extended space-time mentioned earlier, and
 then from this extended manifold to the usual Newtonian 
 space-time \cite{jackiw}.

Parity refers to a reversal of orientation of the spatial manifold. 
 This corresponds to the reversal of coordinates in even-dimensional
 Minkowski space-times.  In odd-dimensional space-times, in which the number
 of spatial coordinates is even, the reflection of spatial manifold
 has determinant equal to one and hence it is continuously connected to the
 identity, and so can be obtained as a rotation.  Therefore, we must 
 define parity as the reversal of sign of an odd number of spatial 
 coordinates in order to reverse the orientation of the spatial
 volume. This is the reason why we start in the $(5+1)$ Minkowski
 space-time in order to define parity operation
 in the $(4+1)$ Galilean space-time.

The development of 8-dimensional gamma matrices for the Dirac equation
 is motivated by applications to problems like the beta decay
 in the 4-fermion Lagrangian of the $V-A$ theory. This requires an
 evaluation of operators like
\[
\psibar_h\gamma_\mu(1-\gamma^5)\psi_h\; \psi_l\gamma^\mu(1-\gamma^5)\psi_l,
\]
which are a combination of the hadron and lepton currents.  Hence the
 necessity to have a $\gamma^5$ matrix which provides us with a chirality
 operator. The leptonic part will be Poincar\'e invariant and the hadronic
 part will be Galilean invariant.  The simplest example is the neutron
 decay:
\[
n\longrightarrow p+e^-+\nubar.
\]
This will provide us with an amplitude that still possesses a symmetry
 instead of just using an expansion in terms of $p/m$, thus destroying
 any symmetry in the hadronic part.

Another application of $\gamma^5$ matrices is in deriving an $N-N$ potential
 with a pseudo-vector or pseudo-scalar coupling.  Although there is no Yukawa
 coupling in Galilean covariant theories, it is still possible to define a 4-point
 coupling.  This will provide an analogue of the $\pi$-meson exchange
 $N-N$ potential in the Poincar\'e-invariant theories.  In addition, it
 is obvious that the interaction term has similarities with the
 Nambu-Jona-Lasinio theory \cite{njl}.  Such a development may also be followed in
 order to obtain further results for the strongly interacting hadronic
 systems. Our purpose is to make progress along these lines with a Galilean covariant
 theory in $(4+1)$ space-time.  However, in order to define the $\gamma^5$-like
 matrix, it is necessary to further extend the theories to a $(5+1)$ 
 Minkowski manifold. Results of this article are therefore quite important
 in order to gain an understanding of the associated physical phenomena.

In section 2, we give an 8-dimensional realization of the Clifford
 algebra in the $(5+1)$ Minkowski space-time.   Then, in section 3,
 we construct wave functions for the Dirac equation in this space-time.
 By dimensional reduction from the $(5+1)$ Minkowski space-time to the
 $(4+1)$ Minkowski space-time, we obtain $8\times 8$ gamma matrices
 obeying the Clifford algebra in the $(4+1)$ Galilean
 space-time in section 4. The construction of wave functions for
 the Dirac-type equation in the $(4+1)$ Galilean
 space-time is performed in section 5.
 The final section contains concluding remarks.

We establish the commutation and anticommutation relations of 
 $8\times 8$ gamma matrices in appendix A, and their trace
 formulas in appendix B.
 Fierz identities are developed in appendix C.
 Finally, in appendix D, we give
 explicit forms of wave functions obtained in sections 3 and 5.
 Throughout this work, we use the natural units,
 in which $\hbar =1$ and $c=1$.

%

% 8-DIM REALIZATION

% 

\section{An 8-dimensional realization of the Clifford algebra
 in the $(5+1)$ Minkowski space-time}

Before we turn to the 8-dimensional realization,
 we begin this section by recalling some useful properties of
 representations of the gamma matrices.
 The $\gamma$-matrices obey the Clifford algebra:
\beq
\label{clifford}\left\{\gamma^{\mu},\gamma^{\nu}\right\}=2g^{\mu\nu},
\eeq
where we choose the metric tensor to be given by
\beq
g_{\mu\nu}={\rm diag}(1,1,1,1,1,-1)=g^{\mu\nu},
\label{2.2}
\eeq
such that
\[
g_{\mu\lambda}\;g^{\lambda\nu}=\delta_{\mu}^{\ \nu}.
\]

If $\gamma^{\mu}$ and $\gamma'^{\mu}$ are two irreducible sets of gamma matrices which satisfy
 the Clifford algebra given in Eq. (\ref{clifford}), then there exists a non-singular matrix $S$ such that
\[
\gamma'^{\mu}=-S^{-1}\gamma^{\mu}S.
\]
By taking this matrix to be equal to
\[
S=\gamma^{0}=-\gamma_0,
\]
where $\left(\gamma^0\right)^2 =-1$ and  
 $\left(\gamma^0\right)^{-1} =-\gamma^0 =\gamma_0$,
 we can define the Hermitian conjugate of $\gamma^{\mu}$
 as follows:
\[
\left(\gamma^{\mu}\right)^{\dagger}= -\left(\gamma^0\right)^{-1}\gamma^{\mu}\gamma^0=
\gamma^{0}\gamma^{\mu}\gamma^{0}.
\]
This reads explicitly as
\[
\begin{array}{l}
\left(\gamma^i\right)^{\dagger}=\gamma^i,\qquad (i=1,2,3),\\
\left(\gamma^a\right)^{\dagger}=\gamma^a,\qquad (a=4,5),\\
\left(\gamma^0\right)^{\dagger}=-\gamma^0 =\left(\gamma^0\right)^{-1}.
\end{array}
\]

The transposition of gamma matrices is defined by taking $S$ to
 be equal to 
\beq
S=C=-\gamma^0 \hat{C},
\label{2.7}\eeq
so that the transpose is obtained as follows: 
\beq
\begin{array}{rcl}
\left(\gamma^{\mu}\right)^T &=&-C^{-1}\gamma^{\mu}C=-C^{-1}\gamma^0 
\left(\gamma^0\gamma^{\mu}\gamma^0\right )\gamma^0 C,\\
&=&C^{-1}(\gamma^0)^{-1}(\gamma^{\mu})^\dagger\gamma^{0}C,\\
&=&{\hat C}^{-1}\left(\gamma^{\mu}\right)^{\dagger}{\hat C}.\end{array}
\label{2.8}\eeq
Note that
\[
\gamma^{\mu}={\hat C}\left(\gamma^{\mu}\right)^{*}{\hat C}^{-1},
\]
where, with $\ast$ denoting the complex conjugation,
\[
{\hat C}^{\dagger}={\hat C}^{-1}=-{\hat C}^{*}.
\]

%  SUBSECTION  8-dim realization

\subsection{An 8-dimensional realization of the Clifford algebra}

In order to obtain an explicit form of $8\times 8$ gamma matrices in
 a 6-dimensional space-time, let us introduce the following nine matrices:
\begin{equation}
\begin{array}{l}
{\bf{\rho}}={\bf{\sigma}}\otimes I\otimes I,\\
{\bf{\pi}}=I\otimes{\bf{\sigma}}\otimes I,\\
{\bf{\Sigma}}=I\otimes I\otimes{\bf{\sigma}},
\end{array}
\label{rhopisigma}
\end{equation}
where ${\bf{\sigma}}$ are the Pauli matrices: 
\[
I=\left (\begin{array}{cc}1&0\\0&1\end{array}\right ),\ \sigma_1 =
\left (\begin{array}{cc}0&1\\1&0\end{array}\right ),\
  \sigma_2 =\left (\begin{array}{cc}0&-\ri\\\ri&0\end{array}\right ),
\ \sigma_3 =\left (\begin{array}{cc}1&0\\0&-1\end{array}\right ).
\]
Then the matrices defined in Eq. (\ref{rhopisigma}) are 
\[
\begin{array}{ccc}
\rho_1 =\left (\begin{array}{cc}\0&\begin{array}{cc}I&0\\0&I\end{array}\\
\begin{array}{cc}I&0\\0&I\end{array}&\0\end{array}\right ),
 & \rho_2 =\left (\begin{array}{cc}\0&\begin{array}{cc}-\ri I&0\\0&-\ri I\end{array}\\
\begin{array}{cc}\ri I&0\\0&\ri I\end{array}&\0\end{array}\right ),
 & \rho_3 =\left (\begin{array}{cc}\begin{array}{cc}I&0\\0&I\end{array}&\0\\
 \0&\begin{array}{cc}-I&0\\0&-I\end{array}\end{array}\right ),\\
\pi_1 =\left (\begin{array}{cc}\begin{array}{cc}0&I\\I&0\end{array}
 &\0\\\0&\begin{array}{cc}0&I\\I&0\end{array}\end{array}\right ),
 & \pi_2 =\left (\begin{array}{cc}\begin{array}{cc}0&-\ri I\\ \ri I&0\end{array}
 & \0\\\0&\begin{array}{cc}0&-\ri I\\ \ri I&0\end{array}\end{array}\right ),
 & \pi_3 =\left (\begin{array}{cc}\begin{array}{cc}I&0\\0&-I\end{array}
 &\0\\\0&\begin{array}{cc}I&0\\0&-I\end{array}\end{array}\right ),\\
{\bf{\Sigma}}=\left (\begin{array}{cc}\begin{array}{cc}{\bf{\sigma}}&0\\0&{\bf{\sigma}}\end{array}
 &\0\\\0&\begin{array}{cc}{\bf{\sigma}}&0\\0&{\bf{\sigma}}\end{array}\end{array}\right ). & &
\end{array}
\]
The following relations hold among these matrices:
\[
\begin{array}{l}
\left[\rho_k ,\Sigma_l\right] =\left[\pi_k ,\Sigma_l\right] =\left[\rho_k ,\pi_l\right] =0,\\
\rho_k\rho_l =\delta_{kl}+\ri\E_{klm}\rho_m,\\
{[\rho_k, \rho_l]} = 2\ri \E_{klm}\rho_m,\\
{\{\rho_k, \rho_l\}} = 2\delta_{kl},\qquad k, l, m = 1, 2, 3,
\end{array}
\]
with similar relations for the $\pi$s and $\Sigma$s.

To complete our construction, we introduce three mutually orthogonal unit
 vectors, ${\bf{m}}$, ${\bf{n}}$, and ${\bf{l}}$, which are utilized to
 express the gamma matrices as follows:
\beq
\begin{array}{l}
{\bf{m}}\cdot{\bf{\rho}}=\ri\gamma^0,\\
\left({\bf{m}}\times{\bf{n}}\right)\cdot{\bf{\rho}} =\gamma^7,\\
\left({\bf{n}}\cdot{\bf{\rho}}\right)\left({\bf{m}}\cdot{\bf{\pi}}\right) =\gamma^4,\\
\left({\bf{n}}\cdot{\bf{\rho}}\right)\left({\bf{l}}\cdot{\bf{\pi}}\right) =\gamma^5,\\
\left({\bf{n}}\cdot{\bf{\rho}}\right)\left({\bf{n}}\cdot{\bf{\pi}}\right){\bf{\Sigma}} ={\bf{\gamma}}.
\end{array}
\label{2.25}\eeq

We can prove that the $\gamma^{\mu}$s given by these equations satisfy the Clifford algebra
 (\ref{clifford}), and that $\gamma^7$ can be cast in the following form:
\[
\gamma^7=\frac{1}{6!}\epsilon_{\mu\nu\lambda\rho\sigma\tau}
\gamma^{\mu}\gamma^{\nu}\gamma^{\lambda}\gamma^{\rho}\gamma^{\sigma}\gamma^{\tau}=
\gamma^1\gamma^2\gamma^3\gamma^4\gamma^5\gamma^0,
\]
with
\[
\epsilon_{012345}=-\epsilon^{012345}=-\epsilon_{123450}=-1.
\]
Commutation and anticommutation relations involving the $\gamma$-matrices in a
 $(5+1)$ Minkowski manifold are given in appendix A.1, and the corresponding
 trace relations are in appendix B.1.

In the next section, we discuss specific definitions of $\bf{m}$, $\bf{n}$
 and $\bf{l}$ which, in turn, lead to a particular representation of the Dirac matrices.

%% SUBSECTION - PAULI REPRESENTATION        

%%

\subsection{The Pauli representation of gamma matrices}

In this section we construct a representation, in which $\ri\gamma^0$ is diagonal, 
 that we shall refer to as  the `Pauli representation' of $\gamma$-matrices.
 It is obtained by choosing
\[
{\bf{m}}=(0,0,1), \qquad{\bf{n}}=(0,1,0),  \qquad{\bf{l}}=(1,0,0).
\]
Therefore, we find
\beq
\begin{array}{l}
\ri\gamma^0 =\rho_3=\sigma_3\otimes I\otimes I =\left(\begin{array}{cc}\begin{array}{cc}I&0\\0&I\end{array}
 &\0\\\0&\begin{array}{cc}-I&0\\0&-I\end{array}\end{array}\right ),\\
\gamma^7 =-\rho_1=-\sigma_1\otimes I\otimes I =\left(\begin{array}{cc}\0&\begin{array}{cc}-I&0\\0&-I\end{array}\\
\begin{array}{cc}-I&0\\0&-I\end{array}&\0\end{array}\right ),\\
\gamma^4 =\rho_2\pi_3=\sigma_2\otimes\sigma_3\otimes I=
 \left(\begin{array}{cc}\0&\begin{array}{cc}-\ri I&0\\0&\ri I\end{array}\\
\begin{array}{cc}\ri I&0\\0&-\ri I\end{array}&\0\end{array}\right ),\\
\gamma^5 =\rho_2\pi_1=\sigma_2\otimes\sigma_1\otimes I =  
\left(\begin{array}{cc}\0&\begin{array}{cc}0&-\ri I\\-\ri I&0\end{array}\\
\begin{array}{cc}0&\ri I\\\ri I&0\end{array}&\0\end{array}\right ),\\
{\bf{\gamma}}=\rho_2\pi_2{\bf{\Sigma}}=\sigma_2\otimes\sigma_2\otimes{\bf{\sigma}}
=\left(\begin{array}{cc}\0&\begin{array}{cc}0&-{\bf{\sigma}}\\{\bf{\sigma}}&0\end{array}\\
\begin{array}{cc}0&{\bf{\sigma}}\\-{\bf{\sigma}}&0\end{array}&\0\end{array}\right ).
\end{array}
\label{2.33}
\eeq
Note that this representation is equivalent to the one described in 
 Refs. \cite{rausch, weyl}.
 We can prove it by choosing the following representation:
\[
{\bf{m}}=(1,0,0),\quad {\bf{n}}=(0,0,1),\quad {\bf{l}}=(0,1,0),
\]
which leads to
\[
\begin{array}{l}
\ri\gamma^0=\rho_1=\Sigma_1^{(3)},\\
\gamma^7=-\rho_2=-\Sigma_2^{(3)},\\
\gamma^4=\rho_3\pi_1=\Sigma_3^{(3)},\\
\gamma^5=\rho_3\pi_2=\Sigma_4^{(3)},\\
\gamma^k=\rho_3\pi_3\Sigma_k=\Sigma_{4+k}^{(3)},\qquad (k=1, 2, 3),
\end{array}
\]
where $\Sigma_a^{(3)}$ ($a=1,\dots, 7$) is in the notation defined
 in Eq. (4.1) of  Ref. \cite{rausch}.

%%  SUBSECTION - INDEPENDENT GAMMA

%

\subsection{Number of independent gamma matrices}

Let $n$ be the dimension of space-time, so that the number of $\Gamma$s is $2^n$.
Since we have
\[
(1+x)^n=\sum^{n}_{k=0}{n\choose k}x^k= \sum^n_{k=0}{}_nC_k\;x^k,
\]
and a completely antisymmetric tensor of rank $k$ has $_nC_k$ independent elements, then
 the number of independent $\Gamma$s is
\[
{}_nC_0+{}_nC_1+\cdots +{}_nC_n=\sum^n_{k=0}{}_nC_k\;(1)^k=(1+1)^n=2^n.
\]
Thus there exists $2^n$ linearly independent matrices:
\[
\Gamma^{(k)}_{\mu_1\cdots\mu_k}=
d^{(k)}_{\mu_1\cdots\mu_k,\nu_1\cdots\nu_k}\gamma^{\nu_1}\cdots\gamma^{\nu_k},
\quad (k=0, 1, \dots, n),
\]
where $d^{(k)}$ are operators, described in Ref. \cite{capri},
 which project out the totally antisymmetric part of a rank-$k$ tensor.

In the case of 6-dimensional Minkowski space-time, we have
 $2^6=64$ independent gamma matrices, which we write as
\[
\begin{array}{l}
\Gamma^{(0)}=\un,\qquad (\un\ {\rm is\ the}\ 8\times 8\ {\rm unit\ matrix)},\\
\Gamma^{(1)}_{\mu}=\gamma_{\mu},\\
\Gamma^{(2)}_{\mu\nu}=\sigma_{\mu\nu}=\frac{1}{2\ri}\left(\gamma_{\mu}\gamma_{\nu}-\gamma_{\nu}\gamma_{\mu}\right),\\
\Gamma^{(3)}_{\mu\nu\lambda}=\sigma_{\mu\nu\lambda}=\frac{1}{3}\left(\gamma_{\mu}\sigma_{\nu\lambda}+
\gamma_{\nu}\sigma_{\lambda\mu}+ \gamma_{\lambda}\sigma_{\mu\nu}\right )=
-\frac{1}{6}\;\epsilon_{\mu\nu\lambda\rho\sigma\tau}\sigma^{\rho\sigma\tau}\gamma^7,\\
\Gamma^{(4)}_{\mu\nu\rho\sigma}=-\frac 12\epsilon_{\mu\nu\rho\sigma\xi\eta}\sigma^{\xi\eta}\gamma^7,\\
\Gamma^{(5)}_{\mu\nu\rho\sigma\lambda}=-\epsilon_{\mu\nu\rho\sigma\lambda\xi}\gamma^\xi\gamma^7,\\
\Gamma^{(6)}_{\mu\nu\rho\sigma\lambda\tau}=-\epsilon_{\mu\nu\rho\sigma\lambda\tau}\gamma^7.
\end{array}
\]

To show properties under the Lorentz transformations, we choose the
 following 64 linearly independent matrices:
\[
\gamma_A=\un, \gamma^7, \gamma_\mu, \ri\gamma^7\gamma_\mu,
 \sigma_{\mu\nu}, \gamma^7\sigma_{\mu\nu}, \sigma_{\mu\nu\lambda},
\]
satisfying
\[
\gamma^A\gamma_A=\un,\qquad ({\rm{no\ summation\ over}}\ A),
\]
\[
{\rm{Tr}}(\gamma_A)=0,\qquad {\rm{if}}\ \gamma_A\neq\un,
\]
as well as
\[
{\rm{Tr}}(\gamma^A\gamma_B)=8\delta^A_{\ B}.
\]
By using the charge conjugation matrix $C$ of Eq. (\ref{2.7}),
 we can separate the $\gamma_A$s into symmetric and antisymmetric
 elements as
\beq
(\gamma_A{C})_{\alpha\beta}=\E_A
 (\gamma_A{C})_{\beta\alpha},
\label{2.53}
\eeq
or, equivalently,
\[
(\gamma_A)_\alpha^{\ \beta}=\E_A ({C}^{-1}\gamma_A{C})^\beta_{\ \alpha},
\]
where
\beq
\E_A=\left\{
\begin{array}{cl}
+1 & {\rm{for}}\ {C}, \gamma^7\sigma_{\mu\nu}{C},
 \sigma_{\lambda\mu\nu}C,\\
-1 & {\rm{for}}\ \gamma^7{C}, \gamma_\mu{C},
 \ri \gamma^7\gamma_\mu{C}, \sigma_{\mu\nu}{C},
\end{array}\right.
\label{2.55}
\eeq
We have used the relation ${C}^\dagger={C}^{-1}={C}^\ast$.

Note that the lowercase indices from the beginning of the Greek alphabet, $\alpha$,
 $\beta$, $\gamma$, etc. denote spinor indices, and the lowercase indices from
 the middle of the alphabet, $\xi$, $\kappa$, $\lambda$, etc. are tensor indices.

%% SUBSECTION PARITY

%

\subsection{Parity}

The parity matrix, denoted $\Pi$, is defined by imposing that the equation of motion
 be invariant under the discrete transformation of space reflection:
\[
x^\mu\rightarrow x'^\mu=(-\bfx,x^4,x^5,x^0).
\] 
Consider the Dirac field, then the requirement reads
\beq
\eta^{-1}\Pi^{\dagger}\eta\gamma^\mu \Pi=\left\{
\begin{array}{l}
-\gamma^\mu,\qquad {\rm for}\ \mu=1, 2, 3,\\
\gamma^\mu,\qquad {\rm for}\ \mu=4, 5, 0,
\end{array}\right.
\label{2.57}
\eeq
where the matrix $\eta$ is defined as
\beq
\eta=\ri\gamma^0.
\label{3.8}
\eeq
Hence the Dirac equation is
 invariant under the space reflection.
 
The parity matrix may be expressed by
\[
\Pi=\gamma^4\gamma^5\gamma^0.
\]

%%

%% SECTION - WAVE FUNCTIONS

%

\section{Construction of wave functions for the Dirac equation in
 the $(5+1)$ Minkowski space-time}

In this section, we obtain the wave functions for the Dirac equation
 of motion in the extended $(5+1)$ Minkowski manifold.
 We adopt the methods of constructing wave functions
 developed by Takahashi \cite{takahashi}, in which the Klein-Gordon
 divisor is diagonalized by using the Lorentz boost. 

The Dirac equation for massive particles with mass $m$ is expressed
 in the form:
\beq
\Lambda(\partial)\psi(x)=0,
\label{3.1}\eeq
where the operator $\Lambda(\pd)$ is given by
\[
\Lambda(\pd)=-(\gamma\cdot\pd+m).
\]
Here, the scalar product is denoted by $A\cdot B$ and defined
 by
\[
A\cdot B=g_{\mu\nu}A^\mu B^\nu=A^iB^i+A^aB^a-A^0B^0,
\]
where the lowercase indices from the beginning of the Latin alphabet,
$a$, $b$ , $c$, etc. take the values 4 and 5, and the lowercase indices from the
middle of the Latin alphabet, $i$, $j$, $k$, etc. run from 1 to 3. 

The adjoint equation to Eq. (\ref{3.1}) is obtained by taking its Hermitian
conjugate:
\[
\psibar(x)\;\Lambda(-\stackrel{\leftarrow}{\pd})=0, 
\]
($\stackrel{\leftarrow}{\pd}$ denotes the left-derivative) with
\[
\psibar(x)=\psi^\dagger(x)\eta.
\]
We assume the existence of a non-singular matrix $\eta$ which satisfies
the relation:
\beq
[\eta\Lambda(\pd)]^\dagger=\eta\Lambda(-\pd).
\label{3.6}
\eeq
This condition is equivalent to requiring the hermiticity of the Lagrangian in the form
\[
{\cal L}(x)=\psibar(x)\Lambda(\pd)\psi(x).
\]
Thus we choose $\eta$ as in Eq. (\ref{3.8}).

The operator $d(\pd)$, reciprocal to the operator $\Lambda(\pd)$
 of Eq. (\ref{3.1}), is defined by
\[
\Lambda(\pd)d(\pd)=d(\pd)\Lambda(\pd)=(\pd^2-m^2)\un .
\]
This reciprocal operator is called the `Klein-Gordon divisor'. It is given by
\[
d(\pd)=-(\gamma\cdot\pd-m).
\]

The Dirac field $\psi(x)$ and its charge-conjugate field $\psi_C(x)$ can be expanded in 
terms of $c$-number wave functions with positive and negative frequencies, represented by
$u_p^{(r)}(x)$ and $v_p^{(r)}(x)$, respectively, and two kinds of creation and 
annihilation operators:
\[
\psi(x) = \sum_r\int d\bp\; d^2p^a\; 
 [ u^{(r)}_p(x)a^{(r)}(\bp,p^a) + v^{(r)}_p(x)b^{(r)\dagger}(\bp,p^a)],
\]
\[
\begin{array}{ccl}
\psi_C(x) & := & {\hat C}\psi^\ast (x),\\
 & = & \sum_r\int d\bp\; d^2p^a\; 
 [u^{(r)}_p(x)b^{(r)}(\bp,p^a) + v^{(r)}_p(x)a^{(r)\dagger}(\bp,p^a)],
\end{array}
\]
where
\[
\{a^{(r)}(\bp, p^a), a^{(r')\dagger}(\bp',p^{\prime a})\} =
 \delta_{rr'}\delta(\bp -\bp')\delta^{(2)}(p^a -p^{\prime a}),
\]
\[
\{b^{(r)}(\bp, p^a), b^{(r')\dagger}(\bp',p^{\prime a})\} =
 \delta_{rr'}\delta(\bp -\bp')\delta^{(2)}(p^a -p^{\prime a}),
\]
and all other commutators of similar type vanish. We use the notation
\[
d^2p^a=dp^4dp^5\quad
{\rm{and}} 
\quad
\delta^{(2)}(p^a-p^{\prime a})= \delta (p^4-p^{\prime 4})\delta(p^5-p^{\prime 5}).
\] 
The function $v^{(r)}_p(x)$ is defined by
\[
v^{(r)}_p(x)= {\hat C} u^{(r)\ast}_p(x).
\]
The charge conjugation matrix ${\hat C}$, defined by Eq. (\ref{2.8}), satisfies
\[
[\eta\Lambda(\pd)]^T=[\eta\Lambda(-\pd)]^\ast={\hat C}^{-1}\eta\Lambda(-\pd){\hat C}.
\]

It is convenient to take the functions $u_p^{(r)}(x)$ to be 
 eigenvectors of the operator $-\ri\pd_\mu$:
\[
-\ri\pd_\mu u_p^{(r)}(x)=p_\mu\; u_p^{(r)}(x)
\]
By substituting the Fourier transform of $u_p^{(r)}(x)$ into this equation,
we find
\[
u_p^{(r)}(x)=f_p(x)\;u^{(r)}(\bp,p^a),
\]
\[
v_p^{(r)}(x)=f_p^\ast (x)\;v^{(r)}(\bp,p^a),
\]
where
\[
f_p(x)=(2\pi)^{-5/2}\;e^{\ri p \cdot x},
\]
and
\[
p^0=\sqrt {\bp \cdot \bp+(p^4)^2+(p^5)^2+m^2}.
\]

By following the prescription developed in chapter 5 of Ref. \cite{takahashi},
 we obtain the orthonormality condition and the closure properties
 in the momentum representation:
\[
\ubar^{(r^\prime)}(\bp, p^a)\ri \gamma^0 u^{(r)}(\bp, p^a)=\delta_{rr^\prime},
\]
\[
\vbar^{(r^\prime)}(\bp, p^a)\ri \gamma^0 v^{(r)}(\bp, p^a)=\delta_{rr^\prime},
\]
\[
\sum_r u_\alpha^{(r)}(\bp, p^a)\ubar^{(r)\beta}(\bp, p^a)= \frac {1}{2p^0} d_\alpha^{\ \beta} (\ri p),
\]
\[
\sum_r v_\alpha^{(r)}(\bp, p^a)\vbar^{(r)\beta}(\bp, p^a)=-\frac {1}{2p^0} d_\alpha^{\ \beta} (-\ri p).
\]

Consider a Lorentz transformation matrix $L(\bp, p^a)$ given by
\beq
L(\bp, p^a)= \sqrt{\frac{p^0+m}{2m}}\;\un-\frac{1}{\sqrt{2m(p^0+m)}}
\; \gamma^0 (\bp\cdot\gamma + p^b\gamma^b).
\label{3.26}
\eeq
Then we have
\beq 
L^{-1}(\bp, p^a)\gamma^{\mu}L(\bp, p^a)=\Lambda^{\mu}_{\ \nu}(\bp, p^a)\gamma^{\nu},
\label{3.27}
\eeq
where
\[
\Lambda^0_{\ \nu} (\bp, p^a)= \left(\frac {p^k}{m}, \frac {p^a}{m}, \frac {p^0}{m}\right),
\]
\[
\Lambda^i_{\ \nu} (\bp, p^a)= \left(g^{ik}+\frac{p^i p^k}{m(p^0+m)}, \frac {p^i p^a}{m(p^0+m)}, \frac {p^i}{m}\right),
\]
\[
\Lambda^b_{\ \nu} (\bp, p^a)= \left(\frac {p^b p^k}{m(p^0+m)}, g^{ba}+ \frac {p^b p^a}{m(p^0+m)}, \frac {p^b}{m}\right).
\]
The transformation coefficients $\Lambda^\mu_{\ \nu}$ satisfy the relation
\[
g_{\mu\nu}\Lambda^\mu_{\ \rho}(\bp, p^a)\Lambda^\nu_{\ \sigma}(\bp, p^a)=g_{\rho\sigma},
\]
as is expected, and hence they induce the homogenous Lorentz transformation. It follows 
from Eq. (\ref {3.27}) that
\beq
L^{-1}(\bp, p^a)d(\ri p)L(\bp, p^a)=m(\un + \ri\gamma^0). 
\label{3.32}
\eeq
The factor $(\un + \ri\gamma^0)$ plays a crucial role when constructing wave functions, because 
we find the following key relations from this factor:
\[
L(\bp, p^a)(\un+\ri\gamma^0)= \frac{1}{\sqrt{2m(p^0+m)}}\;d(\ri p)(\un+\ri\gamma^0),
\]
\[
(\un+\ri\gamma^0)L^{\dagger}(\bp, p^a)\eta=(\un+\ri\gamma^0)L^{-1}(\bp, p^a),
\]
where we have used the relation
\[
\gamma^0 L^{\dagger}(\bp, p^a)\gamma^0=-L^{-1}(\bp, p^a).
\]
Note that Eq. (\ref {2.8}) leads to the useful relation:
\[
{\hat C}L^{\ast}(\bp, p^a){\hat C}^{-1}=L(\bp, p^a).
\]
If we choose the Pauli representation for the gamma matrices, the relation (\ref{3.32}) states that 
the Klein-Gordon divisor is diagonalized by the Lorentz boost (\ref{3.26}).

The helicity operator $h$ is defined in terms of the rank-3 Pauli-Lubanski tensor: 
\beq
h=-\frac{1}{2}\frac{1}{|\bp|}w_{045}=\frac{1}{2}\Sigma_k \frac{p^k}{|\bp|},
\label{3.37}
\eeq 
where $\Sigma_k$ is defined in Eq. (\ref{rhopisigma}), and
 the complete Pauli-Lubanski tensor is given by 
\[
w_{\lambda\mu\nu}=\frac{1}{2}\E_{\lambda\mu\nu\rho\alpha\tau}p^{\rho}\sigma^{\alpha\tau}.
\]
By using the representation of gamma matrices given in Eq. (\ref{2.25}), we find
\[
[L(\bp, p^a),h]=0.
\]
To diagonalize the helicity operator (\ref{3.37}), we introduce the following unitary matrix:
\beq
S(\bp)=\frac{1}{\sqrt{2(1+n^3)}}\;[(1+n^3)\un+ \ri n^2\Sigma_1-\ri n^1 \Sigma_2],
\label{3.40}
\eeq
where
\[
n^k=\frac{p^k}{|\bp|}.
\]
By using the matrix (\ref{3.40}), we can prove the relation
\[
S^{-1}(\bp)h S(\bp)=\frac {1}{2}\Sigma_3,
\]
where $\Sigma_3$ is defined in Eq. (\ref{2.25}),
 hence the helicity operator $h$ is diagonalized by $S(\bp)$. 
The definition in Eq. (\ref{3.40}) implies that
\[
{\hat C}S^{\ast}(\bp){\hat C}^{-1}=S(\bp).
\]

By noticing that
\[
{\hat C}h^{\ast}{\hat C}^{-1}=-h,
\]
we find
\[
h u^{(r)}(\bp, p^a)=\frac{1}{2}\E^{(r)}u^{(r)}(\bp, p^a),
\]
and 
\[
h v^{(r)}(\bp, p^a)=h{\hat C}u^{(r)\ast} (\bp, p^a)=-\frac{1}{2}\E^{(r)}v^{(r)}(\bp, p^a),
\]
in the Pauli representation, where $r$ runs from 1 to 4, and $\E^{(r)}$ is given by
\[
\E^{(r)}=\left\{
\begin{array}{cl}
1 & {\rm{for}}\ r=1,3,\\
-1 & {\rm{for}}\ r=2,4.\end{array}\right.
\]

Wave functions are constructed in the Pauli representation for gamma matrices as follows:
\beq
\begin{array}{rcl}
h_\alpha^{\ \beta} u_\beta^{(r)}(\bp, p^a)&=&\frac{1}{2}\E^{(r)}u_\alpha^{(r)}(\bp, p^a),\\
&=&\frac{1}{\sqrt{4mp^0}}h_\alpha^{\ \beta}\;[d(\ri p)L(\bp, p^a)S(\bp)]_\beta^{\ r},\\
&=&\sqrt{\frac{m}{p^0}}h_\alpha^{\ \beta}\;[L(\bp, p^a)\;\frac{1}{2}(\un+\ri\gamma^0)S(\bp)]_\beta^{\ r},\\
&=&\frac{1}{\sqrt{2p^0(p^0+m)}}\;[(-\ri\gamma\cdot p+m)\;\frac{1}{2}(\un+\ri\gamma^0)\;hS(\bp)]_\alpha^{\ r},\\
&=&\frac{1}{2}\frac{1}{\sqrt{2p^0(p^0+m)}}\;[(-\ri\gamma\cdot p+m)\frac{1}{2}(\un+\ri\gamma^0)S(\bp)\Sigma_3]_\alpha^{\ r},
\end{array}
\label{3.48}
\eeq
\[
\begin{array}{rcl}
\ubar^{(r)\beta}(\bp, p^a)h_\beta^{\ \alpha}&=&\frac{1}{2}\E^{(r)}\ubar^{(r)\alpha}(\bp, p^a),\\
 &=&\sqrt{\frac{m}{p^0}}\;[S^{-1}(\bp)\frac{1}{2}(\un+\ri\gamma^0)L^{-1}(\bp, p^a)]_r^{\ \beta} h_\beta^{\ \alpha},\\
 &=&\frac{1}{2}\;\frac{1}{\sqrt{2p^0(p^0+m)}}\;[\Sigma_3 S^{-1}(\bp)\frac{1}{2}(\un+\ri\gamma^0)
(-\ri\gamma\cdot p+m)]_r^{\ \alpha},
\end{array}
\]
\[
\begin{array}{rcl}
h_\alpha^{\ \beta} v_\beta^{(r)}(\bp, p^a)&=&-\frac{1}{2}\E^{(r)}v_\alpha^{(r)}(\bp, p^a),\\
&=&\sqrt{\frac{m}{p^0}}h_\alpha^{\ \beta}\;[L(\bp, p^a)\; 
\frac{1}{2}(\un-\ri\gamma^0)S{(\bp)}\Sigma_3 {\hat C}]_{\beta r},\\
&=&\frac{1}{2}\frac{1}{\sqrt{2p^0(p^0+m)}}\;[(\ri\gamma\cdot p+m)
\frac{1}{2}(\un-\ri\gamma^0)S(\bp)\Sigma_3 {\hat C}]_{\alpha r},
\end{array}
\]
\beq
\begin{array}{rcl}
\vbar^{(r)\beta}(\bp, p^a)h^{\ \alpha}_\beta&=&-\frac{1}{2}\E^{(r)}\vbar^{(r)\alpha}(\bp, p^a),\\
&=&-\sqrt{\frac{m}{p^0}}\;[{\hat C}^{-1}S^{-1}(\bp)\frac 12(\un-\ri\gamma^0)L^{-1}(\bp, p^a)]^{r\beta}h_\beta^{\ \alpha},\\
&=&-\frac{1}{2}\frac{1}{\sqrt{2p^0(p^0+m)}}\;[{\hat C}^{-1}\Sigma_3 S^{-1}(\bp)\frac{1}{2}(\un-\ri\gamma^0)
(\ri\gamma\cdot p+m)]^{r\alpha},\end{array}
\label{3.51}
\eeq
where the charge conjugation matrix, obtained from Eq. (\ref {2.8}), is given by
\beq
{\hat C}=\gamma^0\gamma^4\gamma^5\gamma^2=\left(\begin{array}{cc}
\0&\begin{array}{cc}-\ri\sigma_2&0\\0&-\ri\sigma_2\end{array}\\
\begin{array}{cc}-\ri\sigma_2&0\\0&-\ri\sigma_2\end{array}&\0\end{array}\right)=-{\hat C}^{-1}.
\label{3.52}
\eeq
To obtain the explicit form of the charge
 conjugation matrix, we have to fix a representation of the gamma matrices.
 We thus found the matrix (\ref{3.52}) in the Pauli representation.
 Explicit forms of wave functions of the form (\ref{3.48}) to (\ref{3.51}) are shown in
 the appendix D.1.

%%

%%  SECTION - 8-DIM REALIZATION IN 4+1 GALILEAN SPACE-TIME

\section{An 8-dimensional realization of the Clifford algebra
 in the 5-dimensional Galilean space-time}

In this section, we turn to the reduction from the $(5+1)$ Minkowski manifold
 to the $(4+1)$ Galilean space-time. More specifically, we exploit 
 the results found in the previous sections to obtain $8\times 8$ gamma
 matrices (denoted $\Gamma$) in the Galilean space-time, from the gamma
 matrices (denoted $\gamma$) defined on the extended Minkowski manifold. 

Consider the 5-dimensional Galilean space-time with light-cone coordinates,
 $x^\mu$ ($\mu=1,\dots,5$), with the metric tensor:
\[
\eta_{\mu\nu}=\left (\begin{array}{cc} {\bf 1}_{3\times 3} & {\bf 0}_{3\times 2}\\
 {\bf 0}_{2\times 3} & \begin{array}{cc} 0 & -1\\ -1 & 0\end{array}\end{array}\right ).
\]
The coordinate system $y^\mu$ ($\mu=1, 2, 3, 4, 0$), defined by
 (see Ref. \cite{omote})
\beq
{\bf y}=\bfx,\quad y^4=\frac 1{\sqrt{2}}(x^4-x^5),\quad
 y^0=\frac 1{\sqrt{2}}(x^4+x^5),
\label{4.1}\eeq
admits the diagonal metric of Eq. (\ref{2.2}). Therefore, the
 5-dimensional Galilean space-time corresponds to a $(4+1)$
 Minkowski space-time, so that it is possible to describe non-relativistic
 theories in a Lorentz-like covariant form. A further reduction, to
 the Newtonian space-time, is needed, as explained
 in Refs. \cite{takahashi1988, omote}. 

In order to introduce pseudo-tensor interactions of rank 0, 1
 and 2 into the 5-dimensional Galilean theory, we need 
 a gamma-6 matrix (which corresponds to the gamma-5 matrix in the
 usual $(3+1)$ Minkowski space-time) obtained by  
  dimensional reduction from
 the $(5+1)$ Minkowski space-time to the $(4+1)$ Minkowski
 space-time with light-cone coordinates.

Let $\Gamma^\mu$ and $\gamma^\mu$ be $8\times 8$ gamma matrices
 in the 5-dimensional Galilean and Minkowski space-times,
 respectively.  They transform as the contravariant vectors in each
 space-time. Therefore, we have
\beq
\begin{array}{l}
\bGamma=\bgamma,\\
\Gamma^4=\frac 1{\sqrt{2}}(\gamma^4+\gamma^0),\\
\Gamma^5=\frac 1{\sqrt{2}}(-\gamma^4+\gamma^0).
\label{4.9}\end{array}
\eeq
The gamma-6 matrix may be taken as
\[
\Gamma^6=\gamma^7,
\]
where $\Gamma^6$ anticommutes with $\Gamma^\mu$.
 Note that neither 
 $\gamma^1\gamma^2\gamma^3\gamma^4\gamma^0$ nor
 $\Gamma^1\Gamma^2\Gamma^3\Gamma^4\Gamma^5$ anticommute
 with the $\Gamma^\mu$s, which satisfy the
 Clifford algebra:
\[
\{\Gamma^\mu,\Gamma^\nu\}=2\eta^{\mu\nu}.
\]

The parity matrix $\Pi$ may be expressed by 
\[ 
\Pi=\gamma^4\gamma^5\gamma^0,
\]
and satisfies the relations
\[
\begin{array}{l}
\Pi\Gamma^k+\Gamma^k\Pi=0,\qquad (k=1,2,3),\\
\Pi\Gamma^4-\Gamma^4\Pi=0,\\
\Pi\Gamma^5-\Gamma^5\Pi=0.
\end{array}
\]
These equations are equivalent to imposing the condition
 given by Eq. (\ref{2.57}). 

Since, in the 5-dimensional Galilean space-time, the dimension of algebra is $2^5=32$,
 then we take thirty-two independent gamma matrices given by
\[
\Gamma_A=\un, \Gamma^6, \Gamma_\mu, \ri\Gamma^6\Gamma_\mu, \Sigma_{\mu\nu}, \Gamma^6\Sigma_{\mu\nu}, 
\]
where $\Sigma_{\mu\nu}$ is defined by
\beq
\Sigma_{\mu\nu}=\frac{1}{2\ri}(\Gamma_\mu\Gamma_\nu-\Gamma_\nu\Gamma_\mu).
\label{4.17}\eeq
These $\Gamma$-matrices satisfy the relation:
\beq
\tr(\Gamma^A\Gamma_B)=8\delta^A_{\ B}.
\label{4.18}
\eeq 
Since the $\Gamma^\mu$s are linear combinations of 
$\gamma^\mu$s and $\Gamma^6=\gamma^7$, we have
\[
(\Gamma^\mu)^T=-{C}^{-1}\Gamma^\mu{C}={\hat C}^{-1}(\Gamma^\mu)^\dagger{\hat C}.
\]
Thus we find
\beq
(\Gamma_A {C})_{\alpha\beta}=\E_A(\Gamma_A{C})_{\beta\alpha}, 
\label{4.20}\eeq
where
\beq
\E_A= \left\{
\begin{array}{cl}
+1 & {\rm{for}}\ {C}, \Gamma^6\Sigma_{\mu\nu}C,\\
-1 & {\rm{for}}\  \Gamma^6{C}, \Gamma_\mu{C},
 \ri\Gamma^6\Gamma_\mu{C}, \Sigma_{\mu\nu}{C}.
\end{array}\right.
\label{4.21}\eeq

%%

%%  SUBSECTION - DIRAC-TYPE EQUATION IN PAULI REPRESENTATION

\subsection{The Dirac-type equation in the Pauli representation}

In the 5-dimensional Galilean space-time, the Dirac-type equation for massless fields 
 can be cast in the following form:
\beq
\Lambda(\partial)\psi(x)=0,
\label{4.22}
\eeq
with
\[
\Lambda(\partial)=-\Gamma^\mu\partial_\mu,
\]
where the wave function is an 8-component spinor. The adjoint equation to
Eq. (\ref{4.22}) is given by
\[
\psibar(x)\Lambda(-\stackrel{\leftarrow}{\pd})=0,
\]
and
\[
\psibar(x)=\psi^\dagger(x)\eta.
\]
Also, we use
\beq
\eta=\ri\frac{1}{\sqrt2}(\Gamma^4+\Gamma^5)=\ri\gamma^0,
\label{4.26}
\eeq
which agrees with Eq. (\ref{3.8}).
Here, we have imposed the relation:
\[
[\eta\Lambda(\pd)]^\dagger=\eta\Lambda(-\pd)\;.
\]

For the fifth component of the derivative $\pd_\mu$, we have the relationship
$\pd_5=-\ri m$, which implies the ansatz
\[
\psi(x)=e^{-\ri m x^5}\psi(\bfx,t),
\]
or, in the matrix form,
\[
\left(\begin{array}{c}
u_1(x)\\
u_2(x)\\
u_3(x)\\
u_4(x)\end{array}\right)=e^{-\ri m x^5}
\left(\begin{array}{c}
u_1(\bfx,t)\\
u_2(\bfx,t)\\
u_3(\bfx,t)\\
u_4(\bfx,t)\end{array}\right)
\]
where $u_k(x)$ and $u_k(\bfx, t)$ ($k=1$, 2, 3, 4) are two-component spinors.

The Galilean $\Gamma$-matrices can be expressed in terms of the $\gamma$-matrices in the
$(5+1)$ Minkowski space-time. By using Eqs. (\ref{2.33}), we obtain the Dirac-type 
equation in the Pauli representation. If we write it out explicitly, we have
\beq
\ri\pd_0[u_1(\bfx, t)+u_3(\bfx, t)]=-\frac{1}{2m}\Delta[u_1(\bfx, t)+u_3(\bfx, t)],
\label{4.30}
\eeq
\[
\ri\pd_0[u_2(\bfx, t)-u_4(\bfx, t)]=-\frac{1}{2m}\Delta[u_2(\bfx, t)-u_4(\bfx, t)],
\]
with
\[
u_1(\bfx, t)-u_3(\bfx, t)=\frac{1}{\sqrt{2}m}\bsig\cdot\bfn [u_2(\bfx, t)-u_4(\bfx, t)]\;,
\]
\beq
u_2(\bfx, t)+u_4(\bfx, t)=\frac{1}{\sqrt{2}m}\bsig\cdot\bfn[u_1(\bfx, t)+u_3(\bfx, t)]\;.
\label{4.33}
\eeq

It is convenient to introduce the orthogonal matrix $R$:
\beq
R=\frac{1}{\sqrt{2}}\left(\begin{array}{cccc}
I&0&I&0\\
0&I&0&I\\
I&0&-I&0\\
0&I&0&-I\end{array}\right)
=\frac{1}{\sqrt{2}}(\rho_1+\rho_3).
\label{4.34}
\eeq
We can utilize this matrix to rotate $\psi(\bfx, t)$ in the form:
\[
\Psi(\bfx, t)=R\psi(\bfx, t).
\]
Written explicitly in matrix form, it reads
\[
\left(\begin{array}{c}
U_1(\bfx, t)\\
U_2(\bfx, t)\\
U_3(\bfx, t)\\
U_4(\bfx, t)\end{array}\right)
=\frac{1}{\sqrt{2}}
\left(\begin{array}{c}
u_1(\bfx, t)+u_3(\bfx, t)\\
u_2(\bfx, t)+u_4(\bfx, t)\\
u_1(\bfx, t)-u_3(\bfx, t)\\
u_2(\bfx, t)-u_4(\bfx, t)\end{array}\right).
\]
Therefore, we obtain from Eqs. (\ref{4.30}) to (\ref{4.33}) that
\beq
\ri\pd_0 U_1(\bfx, t)=-\frac{1}{2m}\Delta\;U_1(\bfx, t)\;,
\label{4.37}\eeq
\[
U_2(\bfx, t)=\frac{1}{\sqrt{2}m}\bsig\cdot\bfn\;U_1 (\bfx, t)\;,
\]
\[
U_3(\bfx, t)=\frac{1}{\sqrt{2}m}\bsig\cdot\bfn\;U_4 (\bfx, t)\;,
\]
\beq
\ri\pd_0 U_4(\bfx, t)=-\frac{1}{2m}\Delta\;U_4(\bfx, t)\;.
\label{4.40}\eeq
This result shows that the 5-dimensional Galilean matrices can be
 obtained by using a similarity transformation which involves
 the orthogonal matrix $R$ .

%%

%%  SUBSECTION - EXPLICIT FORMS OF GAMMA MATRICES

\subsection{Explicit forms of the Galilean gamma matrices}

Consider the Dirac Lagrangian, written as
\[
{\cal L}(x)=\psibar(x)\Lambda(x)\psi(x),
\]
where
\[
\Lambda(\partial)=-\Gamma^\mu\partial_\mu.
\]
The hermiticity of the Lagrangian leads to the condition
 given by Eq. (\ref{3.6}).  This Lagrangian becomes
\beq
{\cal L}(x)=\Psibar(x){\tilde\Lambda}(\pd)\Psi(x),
\label{4.43}\eeq
where $\Psi$ is given by
\[
\Psi(x)=e^{-i m x^5}\Psi(\bfx, t)=e^{-i m x^5}R\psi(\bfx, t),
\]
and ${\tilde\Lambda}$ is defined as
\beq
{\tilde\Lambda}(\pd)=R\Lambda(\pd)R^{-1}.
\label{4.45}
\eeq

Note that
\[
R=R^T=R^{-1},
\]
Therefore, it follows from Eq. (\ref{4.45}) that
\[
{\tilde\Gamma}^\mu=R\Gamma^\mu R^{-1}\;,
\]
\[
{\tilde\eta}=R\eta R^{-1}\;.
\]
The Dirac-type equation is obtained from the Lagrangian
 given by Eq. (\ref{4.43}):
\beq
{\tilde\Lambda}(\pd)\Psi(\bfx, t)=0\;.
\label{4.49}\eeq
If we express the Galilean gamma matrices in the Pauli representation,
then the Dirac-type equation (\ref{4.49}) leads to Eqs. (\ref{4.37}) to
 (\ref{4.40}).

Explicit forms of the Galilean gamma matrices are given by using the Pauli 
representation as follows:
\[
\Gamma^k=\left(\begin{array}{cc} \0 & 
\begin{array}{cc}0&\sigma_k\\ 
-\sigma_k& 0\end{array}\\
\begin{array}{cc}0&-\sigma_k\\ 
\sigma_k& 0\end{array} & \0
\end{array}
\right ),\qquad (k=1, 2, 3),
\]

\[
\Gamma^4=-\sqrt{2}\ri\left(\begin{array}{cc} \0 & 
\begin{array}{cc}0&0\\ 
0&I\end{array}\\
\begin{array}{cc}I&0\\ 
0&0\end{array} & \0
\end{array}
\right ),
\]

\[
\Gamma^5=-\sqrt{2}\ri\left(\begin{array}{cc} \0 & 
\begin{array}{cc}I&0\\ 
0&0\end{array}\\
\begin{array}{cc}0&0\\ 
0&I\end{array} & \0
\end{array}
\right ),
\]

\[
\eta=\left(\begin{array}{cc} \0 & 
\begin{array}{cc}I&0\\ 
0&I\end{array}\\
\begin{array}{cc}I&0\\ 
0&I\end{array} & \0
\end{array}
\right )
=\rho_1.
\]
Moreover, we find
\[
\Gamma^6=\left(
\begin{array}{cc}
\begin{array}{cc}-I&0\\
0&-I\end{array}& \0 \\
\0 & \begin{array}{cc}I&0\\
0&I\end{array} \end{array}
\right )
=-\rho_3\;,
\]
\[
\Pi=\left(\begin{array}{cc} \0 & 
\begin{array}{cc}0&-\ri I\\ 
\ri I&0\end{array}\\
\begin{array}{cc}0&-\ri I\\ 
\ri I&0\end{array} & \0
\end{array}
\right ).
\]
Here we have replaced ${\tilde\Gamma}^\mu, {\tilde\Gamma}^6, {\tilde\eta}$ 
and ${\tilde\Pi}$ by $\Gamma^\mu, \Gamma^6, \eta$ and $\Pi$, respectively. It
should be noted that the matrix $\Gamma^6$ is block diagonal.

The chirality operators may be defined as
\[
\frac{1}{\sqrt{2}}(\un\pm\Gamma^6).
\]
They are block diagonal, and the chiral eigenstates are given by
\[
\psi_{\pm}(x)=\frac{1}{\sqrt{2}}(\un\mp\Gamma^6)\psi(x)\;.
\]

%%

%% SECTION - WAVE FUNCTIONS DIRAC-TYPE EQUATION

%

\section{Construction of wave functions for the Dirac-type equation}

The main advantage of employing a 5-dimensional Galilean covariant theory is that
 we can perform many calculations in a way analogous to the relativistic treatment.
 Indeed, many of our non-relativistic equations have the same form as the
 corresponding equations in relativistic quantum theory, except that they
 are written in a manifestly covariant form on the $(4+1)$ Minkowski
 space-time.

Let $P^\mu$ and $p^\mu$ be contravariant vectors in the 5-dimensional Galilean and 
Minkowski space-times, respectively. Then they are written as
\beq
P^\mu=(\bp, m, E) ,
\label{5.1}\eeq 
\[
p^\mu=(\bp, p^4, p^0),
\]
with
\[
p^0=\frac{1}{\sqrt{2}}(m+E)\;, \;p^4=\frac{1}{\sqrt{2}}(m-E) ,
\]
where we have used Eq. (\ref{4.1}). Moreover, 
 if we impose the conditions
\[
P_\mu P^\mu=p_\mu p^\mu=-\kappa_m^{\ 2},
\]
and
\[
\kappa_m=\sqrt{2}m.
\]
we find
\[
E=\frac{1}{2m}\bp\cdot\bp+m.
\]
We find a similar expression for $p^0$: 
\[
p^0=\sqrt{\bp\cdot\bp+(p^4)^2+\kappa_m\;^2}=
 \frac 1{2\kappa_m}\bp\cdot\bp+\kappa_m.
\]

When we perform the reduction from the 6-dimensional to the 5-dimensional 
Minkowski space-time, the Lorentz boost, Eq. (\ref{3.26}), becomes
\beq
L(\bp, p^4)=\sqrt{\frac{p^0+\kappa_m}{2\kappa_m}}\ \un-
\frac{1}{\sqrt{2\kappa_m(p^0+\kappa_m)}}\;\gamma^0(\bp\cdot{\bf{\gamma}}+p^4\gamma^4)\\
=L^{-1}(-\bp,-p^4).
\label{5.8}
\eeq
 The Galilean transformation matrix is obtained by substituting Eq. (\ref{4.9}) into Eq. (\ref{5.8}):
\beq
L(\bp,p^4)=\frac{1}{\sqrt{2\kappa_m(p^0+\kappa_m)}}
\left [(p^0+\kappa_m)\un-\frac{1}{\sqrt{2}}(\Gamma^4+\Gamma^5)\bp\cdot{\bf{\Gamma}}
-\frac{1}{2}(-\Gamma^4\Gamma^5+\Gamma^5\Gamma^4)p^4\right] =: G(P).
\label{5.9}
\eeq
Hence we find 
\[
G^{-1}(P)\; \Gamma^\mu\; G(P)=Z^\mu_{\ \nu}(P)\Gamma^\nu\;,
\]
where
\[
Z^i_{\ \nu}(P)=\left(\eta^{ik}+\frac{P^i P^k}{m(E+3m)},\frac{2P^i}{E+3m},\frac{(E+m)P^i}{m(E+3m)}\right)\;,
\]
\[
Z^4_{\ \nu}(P)=\left(\frac{2P^k}{E+3m},\frac{4m}{E+3m},\frac{E-m}{E+3m}\right)\;,
\]
\[
Z^5_{\ \nu}(P)=\left(\frac{(E+m)P^k}{m(E+3m)},\frac{E-m}{E+3m},\frac{(E+m)^2}{m(E+3m)}\right)\;.
\]
The transformation coefficients $Z^\mu_{\ \nu}$ lead to 
\[
\eta_{\mu\nu}Z^\mu_{\ \rho}(P)Z^\nu_{\ \sigma}(P)=\eta_{\rho\sigma}.
\]
By noticing that
\[
d(\ri p)=-(\ri\gamma\cdot p-\kappa_m)=-(\ri\Gamma\cdot P-\kappa_m) =: D(\ri P)\;,
\]
we find
\beq
G^{-1}(P)\; D(\ri P)\; G(P)=\kappa_m\left [\un+\ri\frac{1}{\sqrt{2}}(\Gamma^4+\Gamma^5)\right ].
\label{5.16}
\eeq
We can prove the following relations:
\[
G(P)(\un+\eta)=\frac{1}{\sqrt{2m(E+3m)}}\;D(\ri P)\;(\un+\eta),
\]
\[
(\un+\eta)G^\dagger(P)\;\eta=(\un+\eta)G^{-1}(P),
\]
where we have used
\[
\eta G^\dagger(P)\eta=G^{-1}(P),
\]
with $\eta$ defined by Eq. (\ref{4.26}). If we choose the Pauli representation for 
gamma matrices, then Eq. (\ref{5.16}) shows us that the Klein-Gordon divisor in the 5-dimensional
Galilean space-time is diagonalized by the Galilean boost, Eq. (\ref{5.9}).

Following the prescription developed in section 3, we can construct wave functions for the
Dirac-type equation:
\[
-(\ri\gamma\cdot p+\kappa_m)u^{(r)}(\bp,p^4)=-(\ri\Gamma\cdot P+\kappa_m)u^{(r)}(P)=0\;,
\]
where the matrices $\Gamma^\mu$ are given by Eq. (\ref{4.9}), in the
Pauli representation. The wave functions then take the form:
\beq
\begin{array}{rcl}
h_\alpha^{\ \beta} u_\beta^{(r)}(P)&=&\frac{1}{2}\E^{(r)}u_\alpha^{(r)}(P),\\
&=&\frac{1}{2}\;\frac{1}{\sqrt{(E+m)(E+3m)}}
\left [(-\ri\Gamma\cdot P+\kappa_m)\;\frac{1}{2}\;(\un+\eta)S(\bP)\Sigma_3\right ]_\alpha^{\ r},
\end{array}
\label{5.21}\eeq
\[
\begin{array}{rcl}
\ubar^{(r)\beta}(P)h_\beta^{\ \alpha}&=&\frac{1}{2}\E^{(r)}\ubar^{(r)\alpha}(P),\\
&=&\frac{1}{2}\;\frac{1}{\sqrt{(E+m)(E+3m)}}\left [\Sigma_3 S^{-1}(\bP)\;\frac{1}{2}(\un+\eta)
(-\ri\Gamma\cdot P+\kappa_m)\right ]_r^{\ \alpha},
\end{array}
\]
\[
\begin{array}{rcl}
h_\alpha^{\ \beta} v_\beta^{(r)}(P)&=&-\frac{1}{2}\E^{(r)}v_\alpha^{(r)}(P),\\
&=&\frac{1}{2}\;\frac{1}{\sqrt{(E+m)(E+3m)}}\left [(\ri\Gamma\cdot P+\kappa_m)
\;\frac{1}{2}(\un-\eta)S(\bP)\Sigma_3 {\hat C}\right ]_{\alpha r},
\end{array}
\]
\beq
\begin{array}{rcl}
\vbar^{(r)\beta}(P)h_\beta^{\ \alpha}&=&-\frac{1}{2}\E^{(r)}\vbar^{(r)\alpha}(P),\\
&=&-\frac{1}{2}\;\frac{1}{\sqrt{(E+m)(E+3m)}}\left [{\hat C}^{-1}\Sigma_3 S(\bP)
\frac{1}{2}(\un-\eta)(\ri\Gamma\cdot P+\kappa_m)\right ]^{r\alpha},
\end{array}
\label{5.24}\eeq
where the charge conjugation matrix ${\hat C}$ is given by Eq. (\ref{3.52}).
 These wave functions are given explicitly in appendix D.2.

%%

%% SECTION - CONCLUSION

%

\section{Concluding remarks}

The general idea allowing a covariant treatment of non-relativistic
 theories is to perform a dimensional reduction from $(4+1)$ Minkowski space-time.
 However, this prevents the existence of parity operator, since the
 $\gamma^5$-like matrix has no analogue in odd-dimensional space-time.
 Therefore, in this article, we start with a $(5+1)$ space-time.

An 8-dimensional realization of the Clifford algebra in the
 5-dimensional Galilean space-time is obtained by  reduction from the
 6-dimensional to the 5-dimensional Minkowski space-time
 which encompasses Galilean space-time.  The solutions to the
 Dirac-type equation in the 5-dimensional Galilean space-time
 are shown explicitly in the Pauli representation (see appendix D.2).
 The chiral eigenstates are also obtained by rotating the solution
 just mentioned above by means of Eq. (\ref{4.34}).

Consider an inverse Galilean transformation, obtained
 by substituting the direction $(\bp, p^4)$ with $(-\bp, -p^4)$. 
 Then we can derive the Galilean boost from the Lorentz boost,
 Eq. (\ref{5.8}),
\[
L^{-1}(-\bp,-p^4)=L(\bp,p^4)=G(P)\;,
\]
and hence
\[
G(P)\Gamma^\mu G^{-1}(P)=\Gamma^{\prime\nu} Z_\nu^{\ \mu}(P),
\]
where
\[
\Gamma^{\prime\nu}=(-{\bf{\Gamma}},\Gamma^5,\Gamma^4).
\]
It should be mentioned that $\Gamma^4$ and $\Gamma^5$ are interchanged 
by substituting $p^4$ with $-p^4$. Thus, in the massless limit, we
find
\[
\lim_{m\rightarrow 0}Z_\nu^{\ \mu}(P)=
\left(\begin{array}{ccccc}
1&0&0&v^1&0\\
0&1&0&v^2&0\\
0&0&1&v^3&0\\
0&0&0&1&0\\
v^1&v^2&v^3&\frac{1}{2}\bv\cdot{\bv}&1
\end{array}\right),
\]
which is exactly the proper Galilean transformation.

The construction presented in this article allows a definition of
 a parity operator as well as a chirality operator.  
 We have included important
 information related to the Clifford algebra, such as the
 commutation and anticommutation relations, trace formulas and
 the Fierz identities, in the appendices A, B and C, respectively. 
 A reduction to the $(4+1)$ Minkowski space-time
 that encompasses the Galilean space-time is deduced.

Now the necessary developments to treat Galilean covariant
 theories for applications to problems like $\beta$-decay
 and to develop a theory like the Galilean version of the
 Nambu-Jona-Lasinio problem are possible.

\appendix

%%  APPENDIX A - (ANTI-)COMMUTATION RELATIONS

%%

\section{Commutation and anticommutation relations for
 the gamma matrices}

Hereafter, we provide lists of commutation and anticommutation relations
 for the gamma matrices. Section A.1 contains these relations for the
 $8\times 8$ representations discussed in section 2.1, for
 the $(5+1)$ Minkowski space-time. The corresponding relations for
 the 5-dimensional Galilean space-time $\Gamma$-matrices, introduced
 in section 4, are given in section A.2.

%% APPENDIX A.1

\subsection{The $(5+1)$ Minkowski space-time}

The quantities encountered hereafter are defined in section 2.1.
 The matrices $\Sigma^{[\ ][\ ]}$ are described at the end of the
 present section. 

\[
[\gamma^7,\gamma^\mu]=2\gamma^7\gamma^\mu,
\]
\[
[\gamma^7,\ri\gamma^7\gamma^\mu]=2\ri\gamma^\mu,
\]
\[
[\gamma^7,\sigma^{\mu\nu}]=0,
\]
\[
[\gamma^7,\gamma^7\sigma^{\mu\nu}]=0,
\]
\[
[\gamma^7,\sigma^{\lambda\mu\nu}]=2\gamma^7\sigma^{\lambda\mu\nu},
\]
\[
[\gamma^\mu,\gamma^\nu]=2\ri\sigma^{\mu\nu},
\]
\[
[\gamma^\mu, \ri\gamma^7\gamma^\nu]=-\ri\gamma^7\{\gamma^\mu,\gamma^\nu\},
\]
\[
[\gamma^\rho,\sigma^{\mu\nu}]=-2\ri(g^{\rho\mu}\gamma^\nu-g^{\rho\nu}\gamma^\mu),
\]
\[
[\gamma^\lambda,\gamma^7\sigma^{\mu\nu}]=-\gamma^7\{\gamma^\lambda,\sigma^{\mu\nu}\},
\]
\[
[\gamma^\kappa,\sigma^{\lambda\mu\nu}]=-\frac{1}{6}\E^{\lambda\mu\nu\rho\sigma\tau}
\gamma^7\{\gamma^\kappa,\sigma_{\rho\sigma\tau}\},
\]
\[
[\ri\gamma^7\gamma^\mu,\ri\gamma^7\gamma^\nu]=[\gamma^\mu, \gamma^\nu],
\]
\[
[\ri\gamma^7\gamma^\rho,\sigma^{\mu\nu}]=\ri\gamma^7[\gamma^\rho,\sigma^{\mu\nu}],
\]
\[
[\ri\gamma^7\gamma^\lambda,\gamma^7\sigma^{\mu\nu}]=\ri\gamma^7\{\gamma^\lambda,\sigma^{\mu\nu}\},
\]
\[
[\ri\gamma^7\gamma^\rho,\sigma^{\lambda\mu\nu}]=\ri\gamma^7\{\gamma^\rho,\sigma^{\lambda\mu\nu}\},
\]
\[
[\sigma^{\mu\nu},\sigma^{\rho\sigma}]=2\ri(g^{\mu\sigma}\sigma^{\rho\nu}-g^{\mu\rho}\sigma^{\sigma\nu}-
g^{\nu\sigma}\sigma^{\rho\mu}+g^{\nu\rho}\sigma^{\sigma\mu}),
\]
\[
[\sigma^{\mu\nu},\gamma^7\sigma^{\rho\sigma}]=\gamma^7[\sigma^{\mu\nu},\sigma^{\rho\sigma}],
\]
\[
\begin{array}{rcl}
[\sigma^{\rho\sigma},\sigma^{\lambda\mu\nu}]&=&
2\ri[(g^{\lambda\rho}\sigma^{\sigma\mu\nu}-g^{\lambda\sigma}\sigma^{\rho\mu\nu})+
(g^{\mu\rho}\sigma^{\sigma\nu\lambda}-g^{\mu\sigma}\sigma^{\rho\nu\lambda})+\\
& & \qquad\quad+(g^{\nu\rho}\sigma^{\sigma\lambda\mu}-g^{\nu\sigma}\sigma^{\rho\lambda\mu})],\\
&=&-\frac{1}{6}\E^{\lambda\mu\nu\kappa\tau\eta}[\sigma^{\rho\sigma},\sigma_{\kappa\tau\eta}]\;\gamma^7,\\
&=&-\ri\E^{\lambda\mu\nu\kappa\tau\eta}
(g^\rho_{\ \kappa}\sigma^\sigma_{\ \tau\eta}-g^\sigma_{\ \kappa}\sigma^\rho_{\ \tau\eta})\gamma^7,
\end{array}
\]
\[
[\gamma^7\sigma^{\mu\nu},\gamma^7\sigma^{\rho\sigma}]=[\sigma^{\mu\nu},\sigma^{\rho\sigma}],
\]
\[
[\gamma^7\sigma^{\rho\sigma},\sigma^{\lambda\mu\nu}]=\gamma^7\{\sigma^{\rho\sigma},\sigma^{\lambda\mu\nu}\},
\]
\[
[\sigma^{\lambda\mu\nu},\sigma^{\rho\sigma\tau}]=2(\ri\Sigma^{[\lambda\mu\nu][\rho\sigma\tau]}
+\E^{\lambda\mu\nu\rho\sigma\tau}\gamma^7),
\]
\[
\{\gamma^7,\gamma^\mu\}=0\;,
\]
\[
\{\gamma^7,\ri\gamma^7\gamma^\mu\}=0\;,
\]
\[
\{\gamma^7,\sigma^{\mu\nu}\}=2\gamma^7\sigma^{\mu\nu}\;,
\]
\[
\{\gamma^7,\gamma^7\sigma^{\mu\nu}\}=2\sigma^{\mu\nu}\;,
\]
\[
\{\gamma^7,\sigma^{\lambda\mu\nu}\}=0\;,
\]
\[
\{\gamma^\mu,\gamma^\nu\}=2g^{\mu\nu}\;,
\]
\[
\{\gamma^\mu,\ri\gamma^7\gamma^\nu\}=-\ri\gamma^7[\gamma^\mu,\gamma^\nu]\;,
\]
\[
\{\gamma^\lambda,\sigma^{\mu\nu}\}=-\frac{1}{3}\E^{\lambda\mu\nu\rho\sigma\tau}\sigma_{\rho\sigma\tau}\gamma^7=
2\sigma^{\lambda\mu\nu}\;,
\]
\[
\{\gamma^\rho,\gamma^7\sigma^{\mu\nu}\}=-\gamma^7[\gamma^\rho,\sigma^{\mu\nu}]\;,
\]
\[
\{\gamma^\rho,\sigma^{\lambda\mu\nu}\}=2(g^{\rho\lambda}\sigma^{\mu\nu}+g^{\rho\mu}\sigma^{\nu\lambda}
+g^{\rho\nu}\sigma^{\lambda\mu})\;,
\]
\[
\{\ri\gamma^7\gamma^\mu,\ri\gamma^7\gamma^\nu\}=\{\gamma^\mu,\gamma^\nu\}\;,
\]
\[
\{\ri\gamma^7\gamma^\lambda,\sigma^{\mu\nu}\}=\ri\gamma^7\{\gamma^\lambda,\sigma^{\mu\nu}\}\;,
\]
\[
\{\ri\gamma^7\gamma^\rho,\gamma^7\sigma^{\mu\nu}\}=-\ri[\gamma^\rho,\sigma^{\mu\nu}]\;,
\]
\[
\{\ri\gamma^7\gamma^\kappa,\sigma^{\lambda\mu\nu}\}=\ri\gamma^7[\gamma^\kappa,\sigma^{\lambda\mu\nu}]\;,
\]
\[
\{\sigma^{\mu\nu},\sigma^{\rho\sigma}\}=2(\Sigma^{[\mu\nu][\rho\sigma]}
+\ri\E^{\mu\nu\rho\sigma\eta\xi}\sigma_{\eta\xi}\gamma^7)\;,
\]
\[
\{\sigma^{\mu\nu},\gamma^7\sigma^{\rho\sigma}\}=\gamma^7\{\sigma^{\mu\nu},\sigma^{\rho\sigma}\}\;,
\]
\[
\{\sigma^{\rho\sigma},\sigma^{\lambda\mu\nu}\}=2(\Sigma^{[\rho\sigma][\lambda\mu\nu]}-
\E^{\rho\sigma\lambda\mu\nu\eta}\gamma_\eta\gamma^7)\;,
\]
\[
\{\gamma^7\sigma^{\mu\nu},\gamma^7\sigma^{\rho\sigma}\}=\{\sigma^{\mu\nu}, \sigma^{\rho\sigma}\}\;,
\]
\[
\{\gamma^7\sigma^{\rho\sigma},\sigma^{\lambda\mu\nu}\}=\gamma^7[\sigma^{\rho\sigma}\sigma^{\lambda\mu\nu}]\;,
\]
\[
\begin{array}{l}
\{\sigma^{\lambda\mu\nu},\sigma^{\rho\sigma\tau}\}=
\frac{1}{3}[2(g^{\lambda\rho}\sigma^{\sigma\tau}+g^{\lambda\sigma}\sigma^{\tau\rho}+g^{\lambda\tau}\sigma^{\rho\sigma})
\sigma^{\mu\nu}+\\
\quad\quad\qquad +2(g^{\mu\rho}\sigma^{\sigma\tau}+g^{\mu\sigma}\sigma^{\tau\rho}+
g^{\mu\tau}\sigma^{\rho\sigma})\sigma^{\nu\lambda}+\\
\quad\quad\qquad +2(g^{\nu\rho}\sigma^{\sigma\tau}+g^{\nu\sigma}\sigma^{\tau\rho}+
g^{\nu\tau}\sigma^{\rho\sigma})\sigma^{\lambda\mu}]\\
\quad\qquad -\frac{1}{3}\ri[\gamma^7\gamma^\lambda(\E^{\rho\sigma\tau\mu\zeta\xi}\sigma^{\nu}_{\ \zeta\xi}-
\E^{\rho\sigma\tau\nu\zeta\xi}\sigma^\mu_{\ \zeta\xi})+\\
\quad\quad\qquad +\gamma^7\gamma^\mu(\E^{\rho\sigma\tau\nu\zeta\xi}\sigma^{\lambda}_{\ \zeta\xi}-
\E^{\rho\sigma\tau\lambda\zeta\xi}\sigma^\nu_{\ \zeta\xi})+\\
\quad\quad\qquad +\gamma^7\gamma^\nu(\E^{\rho\sigma\tau\lambda\zeta\xi}\sigma^{\mu}_{\ \zeta\xi}
-\E^{\rho\sigma\tau\mu\zeta\xi}\sigma^\lambda_{\ \zeta\xi})].
\end{array}
\]

The matrices $\Sigma^{[\ ][\ ]}$ are defined by:
\[
\Sigma^{[\mu\nu][\rho\sigma]}=g^{\mu\rho}g^{\sigma\nu}-g^{\mu\sigma}g^{\rho\nu}\;,
\]
\[
\Sigma^{[\rho\sigma][\lambda\mu\nu]}=(g^{\rho\lambda}g^{\sigma\mu}-g^{\sigma\lambda}g^{\rho\mu})\gamma^\nu
+(g^{\rho\mu}g^{\sigma\nu}-g^{\sigma\mu}g^{\rho\nu})\gamma^\lambda+(g^{\rho\nu}g^{\sigma\lambda}-
g^{\sigma\nu}g^{\rho\lambda})\gamma^\mu\;,
\]
\[
\begin{array}{l}
\Sigma^{[\lambda\mu\nu][\rho\sigma\tau]}=(g^{\lambda\rho}g^{\mu\sigma}-g^{\lambda\sigma}g^{\mu\rho})\sigma^{\nu\tau}+
(g^{\lambda\sigma}g^{\mu\tau}-g^{\lambda\tau}g^{\mu\sigma})\sigma^{\nu\rho}+
(g^{\lambda\tau}g^{\mu\rho}-g^{\lambda\rho}g^{\mu\tau})\sigma^{\nu\sigma}+\\
\qquad +(g^{\mu\rho}g^{\nu\sigma}-g^{\mu\sigma}g^{\nu\rho})\sigma^{\lambda\tau}+
(g^{\mu\sigma}g^{\nu\tau}-g^{\mu\tau}g^{\nu\sigma})\sigma^{\lambda\rho}+
(g^{\mu\tau}g^{\nu\rho}-g^{\mu\rho}g^{\nu\tau})\sigma^{\lambda\sigma}+\\
\qquad +(g^{\nu\rho}g^{\lambda\sigma}-g^{\nu\sigma}g^{\lambda\rho})\sigma^{\mu\tau}+
(g^{\nu\sigma}g^{\lambda\tau}-g^{\nu\tau}g^{\lambda\sigma})\sigma^{\mu\rho}+
(g^{\nu\tau}g^{\lambda\rho}-g^{\nu\rho}g^{\lambda\tau})\sigma^{\mu\sigma}.
\end{array}
\]

%% APPENDIX A.2

\subsection{The 5-dimensional Galilean space-time}

The quantities described in this appendix are described in section 4. 
 The matrices $\Sigma^{\mu\nu}$ are defined in Eq. (\ref{4.17}). 

\[
[\Gamma^6,\Gamma^\mu]=2\Gamma^6\Gamma^\mu,
\]
\[
[\Gamma^6,\ri\Gamma^6\Gamma^\mu]=2\ri\Gamma^\mu,
\]
\[
[\Gamma^6,\Sigma^{\mu\nu}]=0,
\]
\[
[\Gamma^6,\Gamma^6\Sigma^{\mu\nu}]=0,
\]
\[
[\Gamma^\mu,\Gamma^\nu]=2\ri\Sigma^{\mu\nu},
\]
\[
[\Gamma^\mu,\ri\Gamma^6\Gamma^\nu]=-\ri\Gamma^6\{\Gamma^\mu,\Gamma^\nu\},
\]
\[
[\Gamma^\rho,\Sigma^{\mu\nu}]=-2\ri(\eta^{\rho\mu}\Gamma^\nu-\eta^{\rho\nu}\Gamma^\mu),
\]
\[
[\Gamma^\lambda,\Gamma^6\Sigma^{\mu\nu}]=-\Gamma^6\{\Gamma^\lambda ,\Sigma^{\mu\nu}\},
\]
\[
[\ri\Gamma^6\Gamma^\mu ,\ri\Gamma^6\Gamma^\nu]=[\Gamma^\mu, \Gamma^\nu],
\]
\[
[\ri\Gamma^6\Gamma^\rho ,\Sigma^{\mu\nu}]=\ri\Gamma^6[\Gamma^\rho, \Sigma^{\mu\nu}],
\]
\[
[\ri\Gamma^6\Gamma^\lambda ,\Gamma^6\Sigma^{\mu\nu}]=-\ri\{\Gamma^\lambda, \Sigma^{\mu\nu}\},
\]
\[
[\Sigma^{\mu\nu},\Sigma^{\rho\sigma}]=
2\ri(\eta^{\mu\sigma}\Sigma^{\rho\nu}-\eta^{\mu\rho}\Sigma^{\sigma\nu}-
\eta^{\nu\sigma}\Sigma^{\rho\mu}+\eta^{\nu\rho}\Sigma^{\sigma\mu}),
\]
\[
[\Sigma^{\mu\nu},\Gamma^6\Sigma^{\rho\sigma}]=\Gamma^6[\Sigma^{\mu\nu},\Sigma^{\rho\sigma}],
\]
\[
[\Gamma^6\Sigma^{\mu\nu},\Gamma^6\Sigma^{\rho\sigma}]=[\Sigma^{\mu\nu},\Sigma^{\rho\sigma}],
\]
\[
\{\Gamma^6,\Gamma^\mu\}=0,
\]
\[
\{\Gamma^6,\ri\Gamma^6\Gamma^\mu\}=0,
\]
\[
\{\Gamma^6,\Sigma^{\mu\nu}\}=2\Gamma^6\Sigma^{\mu\nu},
\]
\[
\{\Gamma^6,\Gamma^6\Sigma^{\mu\nu}\}=2\Sigma^{\mu\nu},
\]
\[
\{\Gamma^\mu,\Gamma^\nu\}=2\eta^{\mu\nu},
\]
\[
\{\Gamma^\mu,\ri\Gamma^6\Gamma^\nu\}=-\ri\Gamma^6[\Gamma^\mu,\Gamma^\nu],
\]
\[
\{\Gamma^\lambda,\Sigma^{\mu\nu}\}=\ri(\Gamma^\lambda\Gamma^\nu\Gamma^\mu-\Gamma^\mu\Gamma^\nu\Gamma^\lambda),
\]
\[
\{\Gamma^\rho,\Gamma^6\Sigma^{\mu\nu}\}=-\Gamma^6 [\Gamma^\rho,\Sigma^{\mu\nu}],
\]
\[
\{\ri\Gamma^6\Gamma^\mu ,\ri\Gamma^6\Gamma^\nu\}=\{\Gamma^\mu,\Gamma^\nu\},
\]
\[
\{\ri\Gamma^6\Gamma^\lambda ,\Sigma^{\mu\nu}\}=\ri\Gamma^6\{\Gamma^\lambda,\Sigma^{\mu\nu}\},
\]
\[
\{\ri\Gamma^6\Gamma^\rho ,\Gamma^6\Sigma^{\mu\nu}\}=-\ri[\Gamma^\rho,\Sigma^{\mu\nu}],
\]
\[
\{\Sigma^{\mu\nu},\Sigma^{\rho\sigma}\}=(\Gamma^\mu\Gamma^\nu\Gamma^\sigma\Gamma^\rho+
\Gamma^\rho\Gamma^\sigma\Gamma^\nu\Gamma^\mu)-
2\eta^{\mu\nu}\eta^{\rho\sigma},
\]
\[
\{\Sigma^{\mu\nu},\Gamma^6\Sigma^{\rho\sigma}\}=\Gamma^6\{\Sigma^{\mu\nu},\Sigma^{\rho\sigma}\},
\]
\[
\{\Gamma^6\Sigma^{\mu\nu},\Gamma^6\Sigma^{\rho\sigma}\}=\{\Sigma^{\mu\nu},\Sigma^{\rho\sigma}\}.
\]

%%  APPENDIX B - TRACES OF THE GAMMA MATRICES

%%

\section{Traces of the gamma matrices}

In this appendix, we give lists of traces involving the gamma matrices. The
 $\gamma$-matrices defined in section 2.1 for the $(5+1)$ Minkowski space-time
 are given in section B.1,  and the $\Gamma$-matrices of section 4 for
 the 5-dimensional Galilean space-time are in B.2.

%% APPENDIX B.1

\subsection{The $(5+1)$ Minkowski space-time}

\[
\tr (\gamma_{\mu_1}\cdots\gamma_{\mu_n})=0, \qquad {\rm{for}}\ n\ {\rm{odd}},
\]
\[
\tr (\gamma_\mu\gamma_\nu)=\;8\;g_{\mu\nu},
\]
\[
\tr (\gamma_\mu\gamma_\nu\gamma_\rho\gamma_\sigma)=\;8\;(g_{\mu\nu}g_{\rho\sigma}-
g_{\mu\rho}g_{\nu\sigma}+g_{\mu\sigma}g_{\nu\rho}),
\]
\[
\begin{array}{l}
\tr (\gamma_\lambda\gamma_\mu\gamma_\nu\gamma_\rho\gamma_\sigma\gamma_\tau)
=8[(g_{\lambda\mu}g_{\nu\rho}-g_{\lambda\nu}g_{\mu\rho}+g_{\lambda\rho}g_{\mu\nu})g_{\sigma\tau} 
-g_{\lambda\rho} (g_{\mu\sigma}g_{\nu\tau}-g_{\mu\tau}g_{\nu\sigma})\\
\quad\qquad -(g_{\lambda\mu}g_{\nu\sigma}-g_{\lambda\nu}g_{\mu\sigma}+g_{\lambda\sigma}g_{\mu\nu})g_{\tau\rho}
-g_{\lambda\sigma}(g_{\mu\tau}g_{\nu\rho}-g_{\mu\rho}g_{\nu\tau}) \\
\quad\qquad +(g_{\lambda\mu}g_{\nu\tau}-g_{\lambda\nu}g_{\mu\tau}+g_{\lambda\tau}g_{\mu\nu})g_{\rho\sigma}
-g_{\lambda\tau}(g_{\mu\rho}g_{\nu\sigma}-g_{\mu\sigma}g_{\nu\rho})],
\end{array}
\]
\[
\tr(\gamma^7)=0,
\]
\[
\tr(\gamma^7\gamma_{\mu_{1}}\cdots\gamma_{\mu_{n}})=0, \qquad {\rm{for}}\ n\ {\rm{odd}},
\]
\[
\tr(\gamma^7\gamma_\mu\gamma_\nu)=0,
\]
\[
\tr(\gamma^7\gamma_\mu\gamma_\nu\gamma_\rho\gamma_\sigma)=0,
\]
\[
\tr(\gamma^7\gamma_\lambda\gamma_\mu\gamma_\nu\gamma_\rho\gamma_\sigma\gamma_\tau)=-8\E_{\lambda\mu\nu\rho\sigma\tau},
\]
\[
\tr(\sigma_{\mu\nu})=0,
\]
\[
\tr(\sigma_{\lambda\mu\nu})=0,
\]
\[
\tr(\sigma_{\mu\nu}\sigma_{\rho\sigma})=8(g_{\mu\rho}g_{\nu\sigma}-g_{\mu\sigma}g_{\nu\rho}),
\]
\[
\begin{array}{l}
\tr(\sigma_{\lambda\mu}\sigma_{\nu\rho}\sigma_{\sigma\tau})
=8\ri[(g_{\lambda\tau}g_{\mu\nu}-g_{\lambda\nu}g_{\mu\tau})g_{\rho\sigma}
-(g_{\lambda\sigma}g_{\mu\nu}-g_{\lambda\nu}g_{\mu\sigma})g_{\tau\rho}\\
\qquad\qquad -g_{\lambda\rho}(g_{\mu\sigma}g_{\nu\tau}-g_{\mu\tau}g_{\nu\sigma})
+g_{\mu\rho}(g_{\nu\sigma}g_{\lambda\tau}-g_{\nu\tau}g_{\lambda\sigma})],
\end{array}
\]
\[
\tr(\sigma_{\lambda\mu\nu}\sigma_{\rho\sigma\tau})=
\frac 83\; 
[g_{\lambda\rho}(g_{\mu\sigma}g_{\nu\tau}-g_{\mu\tau}g_{\nu\sigma})
+g_{\lambda\sigma}(g_{\mu\tau}g_{\nu\rho}-g_{\mu\rho}g_{\nu\tau})
+g_{\lambda\tau}(g_{\mu\rho}g_{\nu\sigma}-g_{\mu\sigma}g_{\nu\rho})],
\]
\[
\tr(\gamma^7\sigma_{\lambda\mu\nu}\sigma_{\rho\sigma\tau})=8\E_{\lambda\mu\nu\rho\sigma\tau}\;,
\]
\[
\tr(\gamma^7\sigma_{\lambda\mu}\sigma_{\nu\rho}\sigma_{\sigma\tau})=-8\ri\E_{\lambda\mu\nu\rho\sigma\tau}\;.
\]

%% APPENDIX B.2

\subsection{The 5-dimensional Galilean space-time}

\[
\tr(\Gamma_{\mu_1}\cdots\Gamma_{\mu_n})=0, \qquad {\rm{for}}\ n\ {\rm{odd}},
\]
\[
\tr(\Gamma_\mu\Gamma_\nu)=8\eta_{\mu\nu},
\]
\[
\tr(\Gamma_\mu\Gamma_\nu\Gamma_\rho\Gamma_\sigma)=
8(\eta_{\mu\nu}\eta_{\rho\sigma}-\eta_{\mu\rho}\eta_{\nu\sigma}+\eta_{\mu\sigma}\eta_{\nu\rho}),
\]
\[
\tr(\Gamma^6)=0,
\]
\[
\tr(\Gamma^6\Gamma_{\mu_1}\cdots\Gamma_{\mu_n})=0, \qquad {\rm{for}}\ n\ {\rm{odd}},
\]
\[
\tr(\Gamma^6\Gamma_\mu\Gamma_\nu)=0,
\]
\[
\tr(\Gamma^6\Gamma_\mu\Gamma_\nu\Gamma_\rho\Gamma_\sigma)=0,
\]
\[
\tr(\Sigma_{\mu\nu})=0,
\]
\[
\tr(\Sigma_{\mu\nu}\Sigma_{\rho\sigma})=8(\eta_{\mu\rho}\eta_{\nu\sigma}-\eta_{\mu\sigma}\eta_{\nu\rho}).
\]

%%  APPENDIX C - FIERZ IDENTITIES

%%

\section{Fierz identities}

For later convenience, let us first introduce the following two quantities:
\beq
(\gamma^A,\gamma^B)_{\alpha_{1}\alpha_{2}}^{\ \ \ \ \beta_{1}\beta_{2}}
=\frac{1}{4}[(\gamma^A)_{\alpha_{1}}^{\ \beta_{1}}(\gamma^B)_{\alpha_{2}}^{\ \beta_{2}}+
(\gamma^A)_{\alpha_{1}}^{\ \beta_{2}}(\gamma^B)_{\alpha_{2}}^{\ \beta_{1}}+
(\gamma^B)_{\alpha_{1}}^{\ \beta_{2}}(\gamma^A)_{\alpha_{2}}^{\ \beta_{1}}+
(\gamma^B)_{\alpha_{1}}^{\ \beta_{1}}(\gamma^A)_{\alpha_{2}}^{\ \beta_{2}}],
\label{C.1}
\eeq
\beq
[\gamma^A,\gamma^B]_{\alpha_{1}\alpha_{2}}^{\ \ \ \ \beta_{1}\beta_{2}}   
=\frac{1}{4}[(\gamma^A)_{\alpha_{1}}^{\ \beta_{1}}(\gamma^B)_{\alpha_{2}}^{\ \beta_{2}}
-(\gamma^A)_{\alpha_{1}}^{\ \beta_{2}}(\gamma^B)_{\alpha_{2}}^{\ \beta_{1}}-
(\gamma^B)_{\alpha_{1}}^{\ \beta_{2}}(\gamma^A)_{\alpha_{2}}^{\ \beta_{1}}+
(\gamma^B)_{\alpha_{1}}^{\ \beta_{1}}(\gamma^A)_{\alpha_{2}}^{\ \beta_{2}}].
\label{C.2}
\eeq

From the definition in Eq. (\ref{C.1}), we find the following properties: 
\[
(\gamma^A,\gamma^B)_{\alpha_{1}\alpha_{2}}^{\ \ \ \ \beta_{1}\beta_{2}}
=(\gamma^B,\gamma^A)_{\alpha_{1}\alpha_{2}}^{\ \ \ \ \beta_{1}\beta_{2}},
\]
\[
(\gamma^A+\gamma^B,\gamma^C)_{\alpha_{1}\alpha_{2}}^{\ \ \ \ \beta_{1}\beta_{2}}
=(\gamma^A,\gamma^C)_{\alpha_{1}\alpha_{2}}^{\ \ \ \ \beta_{1}\beta_{2}}+
(\gamma^B,\gamma^C)_{\alpha_{1}\alpha_{2}}^{\ \ \ \ \beta_{1}\beta_{2}},
\]
\[
(\gamma^A,\gamma^B)_{\alpha_{1}\alpha_{2}}^{\ \ \ \ \gamma_{1}\gamma_{2}}
 (\gamma^C,\gamma^D)_{\gamma_{1}\gamma_{2}}^{\ \ \ \ \beta_{1}\beta_{2}}=
\frac 12[ (\gamma^A\gamma^C, \gamma^B\gamma^D)_{\alpha_1\alpha_2}^{\ \ \ \ \beta_1\beta_2} +
 (\gamma^A\gamma^D, \gamma^B\gamma^C)_{\alpha_1\alpha_2}^{\ \ \ \ \beta_1\beta_2} ].
\]
Similarly, from Eq. (\ref{C.2}) we find:
\[
[\gamma^A,\gamma^B]_{\alpha_{1}\alpha_{2}}^{\ \ \ \ \beta_{1}\beta_{2}}
=[\gamma^B,\gamma^A]_{\alpha_{1}\alpha_{2}}^{\ \ \ \ \beta_{1}\beta_{2}},
\]
\[
[\gamma^A+\gamma^B,\gamma^C]_{\alpha_{1}\alpha_{2}}^{\ \ \ \ \beta_{1}\beta_{2}}
=[\gamma^A,\gamma^C]_{\alpha_{1}\alpha_{2}}^{\ \ \ \ \beta_{1}\beta_{2}}+
[\gamma^B,\gamma^C]_{\alpha_{1}\alpha_{2}}^{\ \ \ \ \beta_{1}\beta_{2}},
\]
\[
[\gamma^A,\gamma^B]_{\alpha_{1}\alpha_{2}}^{\ \ \ \ \gamma_{1}\gamma_{2}}
 [\gamma^C,\gamma^D]_{\gamma_{1}\gamma_{2}}^{\ \ \ \ \beta_{1}\beta_{2}}=
\frac 12( [\gamma^A\gamma^C, \gamma^B\gamma^D]_{\alpha_1\alpha_2}^{\ \ \ \ \beta_1\beta_2} -
 [\gamma^A\gamma^D, \gamma^B\gamma^C]_{\alpha_1\alpha_2}^{\ \ \ \ \beta_1\beta_2} ).
\]
Also, the indices admit the following symmetry properties:
\beq
(\gamma^A,\gamma^B)_{\alpha_{1}\alpha_{2}}^{\ \ \ \ \beta_{1}\beta_{2}}=
(\gamma^A,\gamma^B)_{\alpha_{2}\alpha_{1}}^{\ \ \ \ \beta_{1}\beta_{2}}
=(\gamma^A,\gamma^B)_{\alpha_{1}\alpha_{2}}^{\ \ \ \ \beta_{2}\beta_{1}},
\label{C.9}\eeq
\beq
[\gamma^A,\gamma^B]_{\alpha_{1}\alpha_{2}}^{\ \ \ \ \beta_{1}\beta_{2}}=
-[\gamma^A,\gamma^B]_{\alpha_{2}\alpha_{1}}^{\ \ \ \ \beta_{1}\beta_{2}}
=-[\gamma^A,\gamma^B]_{\alpha_{1}\alpha_{2}}^{\ \ \ \ \beta_{2}\beta_{1}}.
\label{C.10}\eeq

The prime Fierz identity comes from the expansion of the identity operator
 $I_{\alpha_{1}\alpha_{2}}^{\ \ \ \ \beta_{1}\beta_{2}}$ in terms of
 $(\gamma^A{C})_{\alpha_{1}\alpha_{2}}$ and $({C}^{-1}\gamma^B)^{\beta_{1}\beta_{2}}$: 
\beq
I_{\alpha_{1}\alpha_{2}}^{\ \ \ \ \beta_{1}\beta_{2}}=
\delta_{\alpha_{1}}^{\ \beta_{1}}\delta_{\alpha_{2}}^{\ \beta_2}
=\sum_{A,B} C_A^{\ \ B}(\gamma^A{C})_{\alpha_{1}\alpha_{2}}({C^{-1}\gamma_B})^{\beta_{1}\beta_{2}}.
\label{C.11}\eeq 
The coefficients $C_A^{\ \ B}$ are determined by using the relation
\beq
(\gamma^A{C})_{\alpha\beta}({C^{-1}\gamma_B})^{\beta\alpha}=\tr(\gamma^A\gamma_B)=8\delta_A^{\ \ B}\;.
\label{C.12}
\eeq
With the operator $\frac{1}{64}({C}^{-1}\gamma_A)^{\alpha_{2}\alpha_{1}}
(\gamma^B{C})_{\beta_{1}\beta_{2}}$ acting on Eq. (\ref{C.11}), we obtain
\[
C_A^{\ \ B}=\frac{1}{64}({C}^{-1}\gamma_A)^{\alpha_{2}\alpha_{1}}
(\gamma^B{C})_{\beta_{1}\beta_{2}}\delta_{\alpha_{1}}^{\ \beta_{1}}
\delta_{\alpha_{2}}^{\ \beta_2}=\frac{1}{8}\delta_A^{\ \ B}\;.
\]
Hence
\beq
\delta_{\alpha_{1}}^{\ \beta_{1}}\delta_{\alpha_{2}}^{\ \beta_2}=
\frac{1}{8}\sum_{A}(\gamma^A{C})_{\alpha_{1}\alpha_{2}}({C^{-1}\gamma_A})^{\beta_{2}\beta_{1}}\;.
\label{C.14}
\eeq
This is the prime Fierz identity in the $(5+1)$ Minkowski space-time.

Since the relationship (\ref{C.12}) holds if we replace $\gamma^A$ with $\Gamma^A$ 
[see Eq. (\ref{4.18})], we can obtain the Fierz identity in the 5-dimensional
Galilean space-time:
\beq
\delta_{\alpha_{1}}^{\ \beta_{1}}\delta_{\alpha_{2}}^{\ \beta_2}=
\frac{1}{8}\sum_{A}(\Gamma^A{C})_{\alpha_{1}\alpha_{2}}({C^{-1}\Gamma_A})^{\beta_{2}\beta_{1}}\;.
\label{C.15}\eeq
It should be noted that the Galilean gamma matrices $\Gamma^A$ are expressed 
in terms of $\gamma^A$s, so that the relationships (\ref{C.1}) to (\ref{C.10}) also hold
for $\Gamma^A$s.

Recalling Eq. (\ref{2.53}) and Eq. (\ref{2.55}), we derive from the prime Fierz identity (\ref{C.14}) that
\beq
\begin{array}{l}
\frac{1}{2}(\delta_{\alpha_{1}}^{\ \beta_{1}}\delta_{\alpha_{2}}^{\ \beta_2}+
\delta_{\alpha_{1}}^{\ \beta_2}\delta_{\alpha_{2}}^{\ \beta_{1}})
=(\un,\un)_{\alpha_{1}\alpha_{2}}^{\ \ \ \beta_{1}\beta_{2}},\\
\quad =\frac{1}{8}[({C})_{\alpha_{1}\alpha_{2}}({C}^{-1})^{\beta_{2}\beta_{1}}
+\frac{1}{2}(\gamma^7\sigma^{\mu\nu}{C})_{\alpha_{1}\alpha_{2}}
({C}^{-1}\gamma^7\sigma_{\mu\nu})^{\beta_{2}\beta_{1}}+\\
\quad +\frac{1}{6}(\sigma^{\lambda\mu\nu}{C})_{\alpha_{1}\alpha_{2}}
({C^{-1}\sigma_{\lambda\mu\nu}})^{\beta_{2}\beta_{1}}],
\end{array}
\label{C.16}\eeq
\beq
\begin{array}{l}
\frac{1}{2}(\delta_{\alpha_{1}}^{\ \beta_{1}}\delta_{\alpha_{2}}^{\ \beta_2}-
\delta_{\alpha_{1}}^{\ \beta_2}\delta_{\alpha_{2}}^{\ \beta_{1}})
=[\un,\un]_{\alpha_{1}\alpha_{2}}^{\ \ \ \beta_{1}\beta_{2}},\\
\quad =\frac{1}{8}[(\gamma^7{C})_{\alpha_{1}\alpha_{2}}({C}^{-1}\gamma^7)^{\beta_{2}\beta_{1}}
 +(\gamma^\mu{C})_{\alpha_{1}\alpha_{2}}({C}^{-1}\gamma_\mu)^{\beta_{2}\beta_{1}}+\\
\quad +(\ri\gamma^7\gamma^\mu{C})_{\alpha_{1}\alpha_{2}}({C}^{-1}\ri\gamma^7\gamma_\mu)^{\beta_{2}\beta_{1}}
+\frac{1}{2}(\sigma^{\mu\nu}{C})_{\alpha_{1}\alpha_{2}}({C^{-1}\sigma_{\mu\nu}})^{\beta_{2}\beta_{1}}].
\end{array}
\label{C.17}\eeq
Similarly, we obtain from Eq. (\ref{C.15}), together with Eqs. (\ref{4.20}) and (\ref{4.21}), that
\beq
\begin{array}{rcl}
\frac{1}{2}(\delta_{\alpha_{1}}^{\ \beta_{1}}\delta_{\alpha_{2}}^{\ \beta_2}+
\delta_{\alpha_{1}}^{\ \beta_2}\delta_{\alpha_{2}}^{\ \beta_{1}})
&=&(\un,\un)_{\alpha_{1}\alpha_{2}}^{\ \ \ \beta_{1}\beta_{2}},\\
&=&\frac{1}{8}[({C})_{\alpha_{1}\alpha_{2}}({C}^{-1})^{\beta_{2}\beta_{1}}
+\frac{1}{2}(\Gamma^6\Sigma^{\mu\nu}{C})_{\alpha_{1}\alpha_(2)}
({C}^{-1}\Gamma^6\Sigma_{\mu\nu})^{\beta_{2}\beta_1}]\;,
\end{array}
\label{C.18}\eeq
\beq
\begin{array}{rcl}
\frac{1}{2}(\delta_{\alpha_{1}}^{\ \beta_{1}}\delta_{\alpha_{2}}^{\ \beta_2}-
\delta_{\alpha_{1}}^{\ \beta_2}\delta_{\alpha_{2}}^{\ \beta_{1}})
&=&[\un,\un]_{\alpha_{1}\alpha_{2}}^{\ \ \ \beta_{1}\beta_{2}},\\
&=&\frac{1}{8}[(\Gamma^6{C})_{\alpha_{1}\alpha_{2}}({C}^{-1}\Gamma^6)^{\beta_{2}\beta_{1}}
+(\Gamma^\mu{C})_{\alpha_{1}\alpha_{2}}({C}^{-1}\Gamma_\mu)^{\beta_{2}\beta_{1}}\\
&+&(\ri\Gamma^6\Gamma^\mu{C})_{\alpha_{1}\alpha_{2}}({C}^{-1}\ri\Gamma^6\Gamma_\mu)^{\beta_{2}\beta_{1}}
+\frac{1}{2}(\Sigma^{\mu\nu}{C})_{\alpha_{1}\alpha_{2}}({C^{-1}\Sigma_{\mu\nu}})^{\beta_{2}\beta_{1}}]\;.
\end{array}
\label{C.19}\eeq

Further Fierz identities follow from Eqs. (\ref{C.16}) and (\ref{C.17}):
\[
\begin{array}{rcl}
(\gamma^A,\gamma^B)_{\alpha_{1}\alpha_{2}}^{\ \ \ \beta_{1}\beta_{2}}&=&
\frac{1}{16}[\E_B(\gamma^A\gamma^B{C})_{\alpha_{1}\alpha_{2}}+
\E_A(\gamma^B\gamma^A{C})_{\alpha_{1}\alpha_{2}}]({C}^{-1})^{\beta_{2}\beta_{1}}\\
&+&\frac{1}{32}[\E_B(\gamma^A\gamma^7\sigma^{\mu\nu}\gamma^B{C})_{\alpha_{1}\alpha_{2}}+
\E_A(\gamma^B\gamma^7\sigma^{\mu\nu}\gamma^A{C})_{\alpha_{1}\alpha_{2}}]
({C}^{-1}\gamma^7\sigma_{\mu\nu})^{\beta_{2}\beta_{1}}\\
&+&\frac{1}{96}[\E_B(\gamma^A\sigma^{\lambda\mu\nu}\gamma^B{C})_{\alpha_{1}\alpha_{2}}
+\E_A(\gamma^B\sigma^{\lambda\mu\nu}\gamma^A{C})_{\alpha_{1}\alpha_{2}}]({C}^{-1}
\sigma_{\lambda\mu\nu})^{\beta_{2}\beta_{1}}\;,
\end{array}
\]
\[
\begin{array}{rcl}
[\gamma^A,\gamma^B]_{\alpha_{1}\alpha_{2}}^{\ \ \ \beta_{1}\beta_{2}}&=&
\frac{1}{16}[\E_B(\gamma^A\gamma^7\gamma^B{C})_{\alpha_{1}\alpha_{2}}
+\E_A(\gamma^B\gamma^7\gamma^A{C})_{\alpha_{1}\alpha_{2}}]({C}^{-1}\gamma^7)^{\beta_{2}\beta_{1}}\\
&+&\frac{1}{16}[\E_B(\gamma^A\gamma^\mu\gamma^B{C})_{\alpha_{1}\alpha_{2}}
+\E_A(\gamma^B\gamma^\mu\gamma^A{C})_{\alpha_{1}\alpha_{2}}]({C}^{-1}\gamma_\mu)^{\beta_{2}\beta_{1}}\\
&+&\frac{1}{16}[\E_B(\gamma^A\ri\gamma^7\gamma^\mu\gamma^B{C})_{\alpha_{1}\alpha_{2}}
+\E_A(\gamma^B\ri\gamma^7\gamma^\mu\gamma^A{C})_{\alpha_{1}\alpha_{2}}]
({C}^{-1}\ri\gamma^7\gamma_\mu)^{\beta_{2}\beta_{1}}\\
&+&\frac{1}{32}[\E_B(\gamma^A\sigma^{\mu\nu}\gamma^B{C})_{\alpha_{1}\alpha_{2}}+
\E_A(\gamma^B\sigma^{\mu\nu}\gamma^A{C})_{\alpha_{1}\alpha_{2}}]
({C}^{-1}\sigma_{\mu\nu})^{\beta_{2}\beta_{1}}\;.
\end{array}
\]
Similarly, from Eqs. (\ref{C.18}) and (\ref{C.19}), we have 
\[
\begin{array}{rcl}
(\Gamma^A,\Gamma^B)_{\alpha_{1}\alpha_{2}}^{\ \ \ \beta_{1}\beta_{2}}&=&
\frac{1}{16}[\E_B(\Gamma^A\Gamma^B{C})_{\alpha_{1}\alpha_{2}}+
\E_A(\Gamma^B\Gamma^A{C})_{\alpha_{1}\alpha_{2}}]({C}^{-1})^{\beta_{2}\beta_{1}}\\
&+&\frac{1}{32}[\E_B(\Gamma^A\Gamma^6\Sigma^{\mu\nu}\Gamma^B{C})_{\alpha_{1}\alpha_{2}}
+\E_A(\Gamma^B\Gamma^6\Sigma^{\mu\nu}\Gamma^A{C})_{\alpha_{1}\alpha_{2}}]
({C}^{-1}\Gamma^6\Sigma_{\mu\nu})^{\beta_{2}\beta_{1}}\;,
\end{array}
\]
\[
\begin{array}{rcl}
[\Gamma^A,\Gamma^B]_{\alpha_{1}\alpha_{2}}^{\ \ \ \beta_{1}\beta_{2}}&=&
\frac{1}{16}[\E_B(\Gamma^A\Gamma^6\Gamma^B{C})_{\alpha_{1}\alpha_{2}}+
\E_A(\Gamma^B\Gamma^6\Gamma^A{C})_{\alpha_{1}\alpha_{2}}]({C}^{-1}\Gamma^6)^{\beta_{2}\beta_{1}}\\
&+&\frac{1}{16}[\E_B(\Gamma^A\Gamma^\mu\Gamma^B{C})_{\alpha_{1}\alpha_{2}}+
\E_A(\Gamma^B\Gamma^\mu\Gamma^A{C})_{\alpha_{1}\alpha_{2}}]({C}^{-1}\Gamma_\mu)^{\beta_{2}\beta_{1}}\\
&+&\frac{1}{16}[\E_B(\Gamma^A\ri\Gamma^6\Gamma^\mu\Gamma^B{C})_{\alpha_{1}\alpha_{2}}+
\E_A(\Gamma^B\ri\Gamma^6\Gamma^\mu\Gamma^A{C})_{\alpha_{1}\alpha_{2}}]
({C}^{-1}\ri\Gamma^6\Gamma_\mu)^{\beta_{2}\beta_{1}}\\
&+&\frac{1}{32}[\E_B(\Gamma^A\Sigma^{\mu\nu}\Gamma^B{C})_{\alpha_{1}\alpha_{2}}
+\E_A(\Gamma^B\Sigma^{\mu\nu}\Gamma^A{C})_{\alpha_{1}\alpha_{2}}]
({C}^{-1}\Sigma_{\mu\nu})^{\beta_{2}\beta_{1}}\;.
\end{array}
\]

%%  APPENDIX D - WAVE FUNCTIONS

%%

\section{Explicit forms of the wave functions}

In this appendix, we display the wave functions explicitly, in both
 the $(5+1)$ Minkowski space-time (appendix D.1) and in the
 $5$-dimensional Galilean space-time (appendix D.2).

%% APPENDIX D.1

\subsection{The $(5+1)$ Minkowski space-time}

The wave functions are given by Eqs. (\ref{3.48}) to
 (\ref{3.51}).  Their explicit expressions are given
 below.

\[
u^{(1)}(\bp,p^a)= \sqrt{\frac{p^0+m}{2p^0}}
\left (
\begin{array}{c}
1\\
0\\
\frac{|\bp|}{p^0+m}n^4\\
\frac{|\bp|}{p^0+m}(\ri+n^5) \end{array}
\right )
\otimes
\frac 1{\sqrt{2(1+n^3)}}\;
\left (
\begin{array}{c}
1+n^3\\
n^1+\ri n^2
\end{array}
\right),
\] 

\[
\ubar^{(1)}(\bp,p^a)=\frac 1{\sqrt{2(1+n^3)}}\;
 (1+n^3, n^1-\ri n^2)\otimes\; \sqrt{\frac{p^0+m}{2p^0}}\;
 \left (1, 0, -\frac{|\bp|}{p^0+m}n^4, -\frac{|\bp|}{p^0+m}(-\ri+n^5)\right ),
\]

\[
u^{(2)}(\bp,p^a)= \sqrt{\frac{p^0+m}{2p^0}}
\left (
\begin{array}{c}
1\\
0\\
\frac{|\bp|}{p^0+m}n^4\\
\frac{|\bp|}{p^0+m}(-\ri+n^5) \end{array}
\right )
\otimes
\frac 1{\sqrt{2(1+n^3)}}\;
\left (
\begin{array}{c}
-n^1+\ri n^2\\
1+n^3
\end{array}
\right),
\] 

\[
\ubar^{(2)}(\bp,p^a)=\frac 1{\sqrt{2(1+n^3)}}\;
 (-n^1-\ri n^2, 1+n^3)\otimes\; \sqrt{\frac{p^0+m}{2p^0}}\;
 \left (1, 0, -\frac{|\bp|}{p^0+m}n^4, -\frac{|\bp|}{p^0+m}(\ri+n^5)\right ),
\]

\[
u^{(3)}(\bp,p^a)= \sqrt{\frac{p^0+m}{2p^0}}
\left (
\begin{array}{c}
0\\
1\\
\frac{|\bp|}{p^0+m}(-\ri+n^5) \\
-\frac{|\bp|}{p^0+m}n^4 \end{array}
\right )
\otimes
\frac 1{\sqrt{2(1+n^3)}}\;
\left (
\begin{array}{c}
1+n^3\\
n^1+\ri n^2
\end{array}
\right),
\] 

\[
\ubar^{(3)}(\bp,p^a)=\frac 1{\sqrt{2(1+n^3)}}\;
 (1+n^3, n^1-\ri n^2)\otimes\; \sqrt{\frac{p^0+m}{2p^0}}\;
 \left (0, 1, -\frac{|\bp|}{p^0+m}(\ri+n^5), \frac{|\bp|}{p^0+m}n^4 \right ),
\]

\[
u^{(4)}(\bp,p^a)= \sqrt{\frac{p^0+m}{2p^0}}
\left (
\begin{array}{c}
0\\
1\\
\frac{|\bp|}{p^0+m}(\ri+n^5) \\
-\frac{|\bp|}{p^0+m}n^4 \end{array}
\right )
\otimes
\frac 1{\sqrt{2(1+n^3)}}\;
\left (
\begin{array}{c}
-n^1+\ri n^2\\
1+n^3
\end{array}
\right),
\] 

\[
\ubar^{(4)}(\bp,p^a)=\frac 1{\sqrt{2(1+n^3)}}\;
 (-n^1-\ri n^2, 1+n^3)\otimes\; \sqrt{\frac{p^0+m}{2p^0}}\;
\left (0, 1, -\frac{|\bp|}{p^0+m}(-\ri+n^5), \frac{|\bp|}{p^0+m}n^4 \right ),
\]

\[
v^{(1)}(\bp,p^a)= \sqrt{\frac{p^0+m}{2p^0}}
\left (
\begin{array}{c}
\frac{|\bp|}{p^0+m} n^4 \\
\frac{|\bp|}{p^0+m} (-\ri + n^5) \\
1\\
0
\end{array}
\right )
\otimes
\frac 1{\sqrt{2(1+n^3)}}\;
\left (
\begin{array}{c}
-n^1+\ri n^2\\
1+n^3
\end{array}
\right),
\] 

\[
\vbar^{(1)}(\bp,p^a)=\frac 1{\sqrt{2(1+n^3)}}\;
 (-n^1-\ri n^2, 1+n^3)\otimes\; \sqrt{\frac{p^0+m}{2p^0}}\;
\left ( \frac{|\bp|}{p^0+m} n^4 , \frac{|\bp|}{p^0+m}(\ri +n^5), -1, 0 \right ),
\]

\[
v^{(2)}(\bp,p^a)= \sqrt{\frac{p^0+m}{2p^0}}
\left (
\begin{array}{c}
-\frac{|\bp|}{p^0+m} n^4 \\
-\frac{|\bp|}{p^0+m}(\ri + n^5) \\
-1\\
0
\end{array}
\right )
\otimes
\frac 1{\sqrt{2(1+n^3)}}\;
\left (
\begin{array}{c}
1+n^3\\
n^1+\ri n^2
\end{array}
\right),
\] 

\[
\vbar^{(2)}(\bp,p^a)=\frac 1{\sqrt{2(1+n^3)}}\;
 (1+n^3, n^1-\ri n^2)\otimes\; \sqrt{\frac{p^0+m}{2p^0}}\;
\left ( -\frac{|\bp|}{p^0+m} n^4, -\frac{|\bp|}{p^0+m}(-\ri + n^5), 1, 0 \right ),
\]

\[
v^{(3)}(\bp,p^a)= \sqrt{\frac{p^0+m}{2p^0}}
\left (
\begin{array}{c}
\frac{|\bp|}{p^0+m}(\ri + n^5) \\
-\frac{|\bp|}{p^0+m} n^4 \\
0\\
 1
\end{array}
\right )
\otimes
\frac 1{\sqrt{2(1+n^3)}}\;
\left (
\begin{array}{c}
-n^1+\ri n^2\\
1+n^3
\end{array}
\right),
\] 

\[
\vbar^{(3)}(\bp,p^a)=\frac 1{\sqrt{2(1+n^3)}}\;
 (-n^1-\ri n^2, 1+n^3)\otimes\; \sqrt{\frac{p^0+m}{2p^0}}\;
\left (\frac{|\bp|}{p^0+m}(-\ri + n^5), -\frac{|\bp|}{p^0+m} n^4, 0, -1 \right ),
\]

\[
v^{(4)}(\bp,p^a)= \sqrt{\frac{p^0+m}{2p^0}}
\left (
\begin{array}{c}
-\frac{|\bp|}{p^0+m}(-\ri + n^5) \\
 \frac{|\bp|}{p^0+m} n^4 \\
0 \\
 -1
\end{array}
\right )
\otimes
\frac 1{\sqrt{2(1+n^3)}}\;
\left (
\begin{array}{c}
1+n^3\\
n^1+\ri n^2
\end{array}
\right),
\] 

\[
\vbar^{(4)}(\bp,p^a)=\frac 1{\sqrt{2(1+n^3)}}\;
 (1+n^3, n^1-\ri n^2)\otimes\; \sqrt{\frac{p^0+m}{2p^0}}\;
\left (-\frac{|\bp|}{p^0+m}(\ri + n^5), \frac{|\bp|}{p^0+m} n^4, 0, 1 \right ),
\]
where we have used the notation
\[
n^a=\frac{p^a}{|\bp|},\qquad (a=4,5).
\]

%% APPENDIX D.2

\subsection{The 5-dimensional Galilean space-time}

In this subsection, we give explicit forms of the wave functions
 given by Eqs. (\ref{5.21}) to (\ref{5.24}). The symbol $P$ is
 a shorthand notation for the 5-momentum defined in Eq.
 (\ref{5.1}).

\[
u^{(1)}(P)= \frac 1{\sqrt{2}}\sqrt{\frac{E+3m}{E+m}}
\left (
\begin{array}{c}
1\\
0\\
- \frac{E-m}{E+3m} \\
\ri 2\frac{\sqrt{m(E-m)}}{E+3m} \end{array}
\right )
\otimes
\frac 1{\sqrt{2(1+n^3)}}\;
\left (
\begin{array}{c}
1+n^3\\
n^1+\ri n^2
\end{array}
\right),
\] 

\[
\ubar^{(1)}(P)=\frac 1{\sqrt{2(1+n^3)}}\;
 (1+n^3, n^1-\ri n^2)\otimes\; \frac 1{\sqrt{2}}\sqrt{\frac{E+3m}{E+m}}
 \left (1, 0, \frac{E-m}{E+3m} , \ri 2\frac{\sqrt{m(E-m)}}{E+3m} \right ),
\]

\[
u^{(2)}(P)= \frac 1{\sqrt{2}}\sqrt{\frac{E+3m}{E+m}}
\left (
\begin{array}{c}
1\\
0\\
- \frac{E-m}{E+3m} \\
- \ri 2\frac{\sqrt{m(E-m)}}{E+3m} \end{array}
\right )
\otimes
\frac 1{\sqrt{2(1+n^3)}}\;
\left (
\begin{array}{c}
-n^1+\ri n^2\\
1+n^3
\end{array}
\right),
\] 

\[
\ubar^{(2)}(P)=\frac 1{\sqrt{2(1+n^3)}}\;
 (-n^1-\ri n^2, 1+n^3)\otimes\; \frac 1{\sqrt{2}}\sqrt{\frac{E+3m}{E+m}}
 \left (1, 0, \frac{E-m}{E+3m} , -\ri 2\frac{\sqrt{m(E-m)}}{E+3m} \right ),
\]

\[
u^{(3)}(P)= \frac 1{\sqrt{2}}\sqrt{\frac{E+3m}{E+m}}
\left (
\begin{array}{c}
0\\
1\\
-\ri 2\frac{\sqrt{m(E-m)}}{E+3m}\\
 \frac{E-m}{E+3m} 
 \end{array}
\right )
\otimes
\frac 1{\sqrt{2(1+n^3)}}\;
\left (
\begin{array}{c}
1+n^3\\
n^1+\ri n^2
\end{array}
\right),
\] 

\[
\ubar^{(3)}(P)=\frac 1{\sqrt{2(1+n^3)}}\;
 (1+n^3, n^1-\ri n^2)\otimes\; \frac 1{\sqrt{2}}\sqrt{\frac{E+3m}{E+m}}
 \left (0, 1, -\ri 2\frac{\sqrt{m(E-m)}}{E+3m}, - \frac{E-m}{E+3m}  \right ),
\]

\[
u^{(4)}(P)= \frac 1{\sqrt{2}}\sqrt{\frac{E+3m}{E+m}}
\left (
\begin{array}{c}
0\\
1\\
\ri 2\frac{\sqrt{m(E-m)}}{E+3m}\\
 \frac{E-m}{E+3m} 
 \end{array}
\right )
\otimes
\frac 1{\sqrt{2(1+n^3)}}\;
\left (
\begin{array}{c}
-n^1+\ri n^2\\
1+n^3
\end{array}
\right),
\] 

\[
\ubar^{(4)}(P)=\frac 1{\sqrt{2(1+n^3)}}\;
 (-n^1-\ri n^2, 1+n^3)\otimes\; \frac 1{\sqrt{2}}\sqrt{\frac{E+3m}{E+m}}
\left (0, 1, \ri 2\frac{\sqrt{m(E-m)}}{E+3m}, - \frac{E-m}{E+3m}  \right ),
\]

\[
v^{(1)}(P)= \frac 1{\sqrt{2}}\sqrt{\frac{E+3m}{E+m}}
\left (
\begin{array}{c}
 -\frac{E-m}{E+3m} \\
-2\ri\frac{\sqrt{m(E-m)}}{E+3m}\\
1\\
0
 \end{array}
\right )
\otimes
\frac 1{\sqrt{2(1+n^3)}}\;
\left (
\begin{array}{c}
-n^1+\ri n^2\\
1+n^3
\end{array}
\right),
\] 

\[
\vbar^{(1)}(P)=\frac 1{\sqrt{2(1+n^3)}}\;
 (-n^1-\ri n^2, 1+n^3)\otimes\; \frac 1{\sqrt{2}}\sqrt{\frac{E+3m}{E+m}}
\left (-\frac{E-m}{E+3m}, 2\ri\frac{\sqrt{m(E-m)}}{E+3m},  -1, 0 \right ),
\]

\[
v^{(2)}(P)= \frac 1{\sqrt{2}}\sqrt{\frac{E+3m}{E+m}}
\left (
\begin{array}{c}
 \frac{E-m}{E+3m} \\
-2\ri\frac{\sqrt{m(E-m)}}{E+3m}\\
-1\\
 0
 \end{array}
\right )
\otimes
\frac 1{\sqrt{2(1+n^3)}}\;
\left (
\begin{array}{c}
1+n^3\\
n^1+\ri n^2
\end{array}
\right),
\] 

\[
\vbar^{(2)}(P)=\frac 1{\sqrt{2(1+n^3)}}\;
 (1+n^3, n^1-\ri n^2)\otimes\; \frac 1{\sqrt{2}}\sqrt{\frac{E+3m}{E+m}}
\left ( \frac{E-m}{E+3m}, 2\ri \frac{\sqrt{m(E-m)}}{E+3m},  1, 0  \right ),
\]

\[
v^{(3)}(P)= \frac 1{\sqrt{2}}\sqrt{\frac{E+3m}{E+m}}
\left (
\begin{array}{c}
2\ri\frac{\sqrt{m(E-m)}}{E+3m}\\ 
 \frac{E-m}{E+3m} \\
0\\
 1
 \end{array}
\right )
\otimes
\frac 1{\sqrt{2(1+n^3)}}\;
\left (
\begin{array}{c}
-n^1+\ri n^2\\
1+n^3
\end{array}
\right),
\] 

\[
\vbar^{(3)}(P)=\frac 1{\sqrt{2(1+n^3)}}\;
 (-n^1-\ri n^2, 1+n^3)\otimes\; \frac 1{\sqrt{2}}\sqrt{\frac{E+3m}{E+m}}
\left ( -2\ri\frac{\sqrt{m(E-m)}}{E+3m}, \frac{E-m}{E+3m}, 0, -1  \right ),
\]

\[
v^{(4)}(P)= \frac 1{\sqrt{2}}\sqrt{\frac{E+3m}{E+m}}
\left (
\begin{array}{c}
 2\ri\frac{\sqrt{m(E-m)}}{E+3m}\\ 
- \frac{E-m}{E+3m} \\
0\\
 -1
 \end{array}
\right )
\otimes
\frac 1{\sqrt{2(1+n^3)}}\;
\left (
\begin{array}{c}
1+n^3\\
n^1+\ri n^2
\end{array}
\right),
\] 

\[
\vbar^{(4)}(P)=\frac 1{\sqrt{2(1+n^3)}}\;
 (1+n^3, n^1-\ri n^2)\otimes\; \frac 1{\sqrt{2}}\sqrt{\frac{E+3m}{E+m}}
\left ( -2\ri\frac{\sqrt{m(E-m)}}{E+3m}, -\frac{E-m}{E+3m}, 0, 1  \right ).
\]

It is important to remark that the solutions in subsection D have well-defined 
 massless limits.

%  ACKNOWLEDGEMENT

\section*{Acknowledgement}

We acknowledge partial support by the Natural Sciences and Engineering
 Research  Council of Canada. This manuscript was completed while M.K. was
 visiting the Theoretical Physics Institute at the University of Alberta.
 The authors wish to thank Profs. R. Jackiw and V.P. Nair for helpful
 comments and suggestions at the early stage of this work.

%   BIBLIOGRAPHY


\begin{thebibliography}{99}
\bibitem{takahashi1988} Takahashi Y \FP{Towards the many-body theory with the
 Galilei invariance as a guide I}{1988}{36}{63-81}

Takahashi Y \FP{Towards the many-body theory with the
 Galilei invariance as a guide II}{1988}{36}{83-96}
\end{thebibliography}
\end{document}